\documentclass[numberedappendix]{emulateapj}
\pdfoutput=1
\usepackage{natbib}
\usepackage{amsmath}
\usepackage{color}
\usepackage{booktabs}
\usepackage{threeparttable}
\usepackage{url}
\definecolor{red}{rgb}{1,0,0}

\newcommand{\lya}{Ly$\alpha$}
\newcommand{\elya}{\mathrm{Ly}\alpha}
\newcommand{\fesc}{f_{\mathrm{esc}}}

\newcommand{\lint}{L_{\elya,\mathrm{intr}}}
\newcommand{\ha}{H$\alpha$}
\newcommand{\hb}{H$\beta$}

\newcommand{\hii}{H{\small II}}
\newcommand{\oii}{[O~{\small II}]}

\newcommand{\oiii}{[O~{\small III}]}
\newcommand{\heii}{He{\small II}}

\newcommand{\nii}{[N~{\small II]}}

\newcommand{\ecs}{erg s$^{-1}$ cm$^{-2}$}
\newcommand{\es}{erg s$^{-1}$}

\newcommand{\se}{\textit{Source Extractor}}
\newcommand{\vband}{V_{606}}
\newcommand{\iband}{I_{814}}
\newcommand{\jband}{J_{110}}
\newcommand{\hband}{H_{160}}
\newcommand\numberthis{\addtocounter{equation}{1}\tag{\theequation}}

\shortauthors{Bagley et al.}

\begin{document}


\title{A high space density of luminous Lyman Alpha Emitters at $\lowercase{z}\sim6.5$}


\author{Micaela B. Bagley\altaffilmark{1},
Claudia Scarlata\altaffilmark{1},
Alaina Henry\altaffilmark{2},
Marc Rafelski\altaffilmark{2},
Matthew Malkan\altaffilmark{3},
Harry Teplitz\altaffilmark{4},
Y. Sophia Dai\altaffilmark{4},
Ivano Baronchelli\altaffilmark{4},
James Colbert\altaffilmark{4},
Michael Rutkowski\altaffilmark{5},
Vihang Mehta\altaffilmark{1},
Alan Dressler\altaffilmark{6},
Patrick McCarthy\altaffilmark{6},
Andrew Bunker\altaffilmark{7},
Hakim Atek\altaffilmark{8},
Thibault Garel\altaffilmark{9},
Crystal L. Martin\altaffilmark{10},
Nimish Hathi\altaffilmark{11,}\altaffilmark{2},
Brian Siana\altaffilmark{12}
}

\altaffiltext{1}{Minnesota Institute for Astrophysics, University of Minnesota, Minneapolis, MN 55455, USA}
\altaffiltext{2}{Space Telescope Science Institute, Baltimore, MD 21218, USA}
\altaffiltext{3}{Department of Physics \& Astronomy, University of California, Los Angeles, CA 90095, USA}
\altaffiltext{4}{Infrared Processing and Analysis Center, California Institute of Technology, Pasadena, CA 91125, USA}
\altaffiltext{5}{Stockholm University, Department of Astronomy and Oskar Klein Centre for Cosmoparticle Physics, AlbaNova University Centre, SE-10691, Stockholm, Sweden}
\altaffiltext{6}{Carnegie Observatories, Pasadena, CA 91101, USA}
\altaffiltext{7}{Department of Physics, University of Oxford, Oxford, UK}
\altaffiltext{8}{Laboratoire d’Astrophysique, Ecole Polytechnique F\'ed\'erale de Lausanne, Observatoire de Sauverny, CH-1290 Versoix, Switzerland}
\altaffiltext{9}{Univ Lyon, Univ Lyon1, Ens de Lyon, CNRS, Centre de Recherche Astrophysique de Lyon UMR5574, F-69230, Saint-Genis-Laval, France}
\altaffiltext{10}{Department of Physics, University of California, Santa Barbara, CA 93106, USA}
\altaffiltext{11}{Laboratoire d’Astrophysique de Marseille, Marseille, France}
\altaffiltext{12}{Department of Physics, University of California, Riverside, CA 92521, USA}

\begin{abstract}
We present the results of a systematic search for Lyman-alpha emitters (LAEs) 
at $6 \lesssim z \lesssim 7.6$ using the \textit{HST} WFC3 Infrared 
Spectroscopic Parallel (WISP) Survey. 
Our total volume over this redshift range is $\sim 8 \times10^5$ Mpc$^3$, 
comparable to many of the narrowband surveys despite their larger 
area coverage. We find two LAEs at $z=6.38$ and $6.44$ 
with line luminosities of L$_{\mathrm{Ly}\alpha} \sim 4.7 \times 10^{43}$ \es, 
putting them among the brightest LAEs discovered at these redshifts.
Taking advantage of the broad spectral coverage of WISP, 
we are able to rule out almost all lower-redshift contaminants.
The WISP LAEs have a high number density of $7.7\times10^{-6}$ Mpc$^{-3}$.
We argue that the LAEs reside in Mpc-scale ionized bubbles that allow the
\lya\ photons to redshift out of resonance before encountering the neutral IGM.
We discuss possible ionizing sources and conclude that the observed LAEs 
alone are not sufficient to ionize the bubbles.
\end{abstract}


\section{Introduction}\label{sec:introduction}
The dark ages that followed recombination ended with the appearance of
metal-free stars and the subsequent formation of numerous
low-mass, metal-poor galaxies.
The collective ionizing background from these newly-forming
galaxies is thought to be responsible for the reionization of the
diffuse hydrogen in the intergalactic medium (IGM)
between $6.5\lesssim z \lesssim
10$ \citep[e.g.,][]{fan2006,robertson2012,madau2015,planck2016}.
The progression of the reionization history depends on the
nature of these first sources -- their number densities, luminosities,
clustering, and production rates of ionizing photons -- 
which is currently the subject of considerable observational and 
theoretical efforts.

The evolving neutral hydrogen fraction in the IGM can be constrained using
observations of the \lya\ output from galaxies around the end of
reionization. \lya\ photons are produced by a resonant transition and so 
are sensitive to the presence of even a small fraction of neutral hydrogen.
In particular, a drop in both the number density of \lya\ emitters and in
the fraction of \lya\ emitters among star-forming galaxies is expected at 
$z\gtrsim 6$ due to the increase in neutral hydrogen in the IGM.

Observational studies comparing the $z\simeq5.7$, $z\simeq6.5$, and
$z\geq7$ \lya\ luminosity functions (LFs) have so far reached inconsistent 
conclusions.
Some find a deficit with increasing redshift in the number density of 
\lya~emitters (LAEs) either at the bright end of the LF 
\citep[e.g.,][]{kashikawa2011}, the faint end of the LF 
\citep[e.g.,][]{matthee2014,matthee2015}, or at all luminosities 
\citep{konno2014}. 
The differences in completeness associated with each survey could contribute to 
this disagreement. For example, the large discrepency between the faint 
ends of the \lya\ LFs of \cite{kashikawa2011} and \cite{hu2010} is likely due to
the former's deeper spectroscopic observations.
Cosmic variance may also contribute to the discrepencies at the bright end.
Recent and ongoing narrowband surveys are covering much larger areas
than previous surveys. 
For example, \cite{matthee2015} and \cite{santos2016} present 
the results of their $7$ sq. degree survey while
\cite{hu2016} report on findings in $3$ sq. degrees of their 
$24$ sq. degree survey. On the other hand, 
\cite{hu2010} covered just $1.16$ sq. degrees, and \cite{kashikawa2011}
went very deep in a single pointing of $\sim0.25$ sq. degrees. 
It is possible that the older surveys, more susceptible to cosmic variance
due to their narrow redshift windows, 
did not cover volumes large enough to probe the true variation in the number 
densities of these rare, brightest objects. 

While observational issues do exist, the observed disagreement could also be 
caused by real astrophysical phenomena. Reionization is not
expected to proceed at the same rate on all scales 
\citep[e.g.,][]{mesinger2008,zheng2011,treu2012},
but rather to be a patchy and inhomogeneous process depending on the
luminosity of ionizing sources and the galaxy clustering
properties. The effect of both luminosity and clustering is to
produce large ionized bubbles (on scales of a few megaparsecs, depending on
the exact luminosity) that allow \lya\ photons to redshift out
of resonance before encountering the IGM.
Therefore, we may expect the most luminous LAEs to be visible out to 
earlier times. Conversely, it will take the fainter galaxies longer to 
ionize bubbles large enough to allow \lya\ photons to escape
\citep[e.g.,][]{matthee2015,ono2012,stark2016}. 
Additionally, around luminous sources and/or highly clustered regions, we may
expect an enhancement in the number density of galaxies showing \lya\ 
in emission \citep{castellano2016}.

Small sample sizes and contamination by lower-redshift interlopers 
are both major concerns in high-redshift LAE searches.  
\cite{tilvi2010} and \cite{krug2012} claim to find evidence that
the number density of LAEs either does not evolve or slightly increases
from $z=5.7-7.7$. These results were likely due to contamination. 
Indeed, the follow-up observations of \cite{faisst2014} find no
\lya\ emission in the spectrum of either LAE candidate from \cite{krug2012}. 
Due to the challenges of performing ground-based narrowband searches 
at $z>7$, the majority of candidates targeted for spectroscopic confirmation
are identified from broadband colors indicative of a Lyman break 
\citep[e.g.,][]{vanzella2011,pentericci2011,pentericci2014,ono2012,schenker2012}.
Such ``dropout'' samples, however, can suffer from large contamination 
fractions, especially for fainter galaxies
\citep[e.g.,][]{dickinson2004,stanway2008}.
Moreover, as surveys for ultra-faint \lya\ emitters have shown 
\citep{dressler2011,dressler2015,henry2012}, 
continuum-based searches may miss significant fractions of faint 
galaxies that are important for reionizing the IGM.

Spectroscopic surveys with \textit{HST}'s  WFC3/IR grism are well suited to 
address the still-uncertain evolution of the bright end of the \lya\ LF 
at $z\sim6-7$.
The WFC3 Infrared Spectroscopic Parallel
(WISP, PI: M. Malkam, \citealt{atek2010}) Survey 
covers more than 1700 sq. arcmin in 386 uncorrelated fields.
Here we present the results of a search for
$z\sim 6-7$ galaxies in the $48$ deepest WISP fields available, covering 
$\sim160$ sq. arcmin.
We compensate for the small area by covering a broad redshift range
($6\lesssim z \lesssim7.6$)
enabling us to probe a volume of $8\times10^5$ Mpc$^3$ at $z>6$. 
We are also able to rule out
almost all lower-redshift contaminants thanks to the broad spectral 
coverage of the WISP survey. 

This paper is organized as follows: in Section \ref{sec:data} we describe 
our WISP observations and data reduction.
We present the selection of $z>6$ LAEs in Section \ref{sec:sample} and 
the expected contamination fraction in Section \ref{sec:contam}. We
present our results in Section \ref{sec:results} and discuss the 
implications in Section \ref{sec:discussion}.
Throughout this paper we assume a cosmology with
$\Omega_0=0.3$, $\Omega_{\Lambda}=0.7$, and
$H_0=70$ km s$^{-1}$ Mpc$^{-1}$.
All magnitudes are expressed in the AB system \citep{oke1983}.

\section{Observations - The WISP Survey}\label{sec:data}
The WISP Survey (PI: M. Malkan, \citealp{atek2010}) is a near-infrared
slitless grism spectroscopic program which efficiently accrues 
WFC3\footnote{\url{http://ww.stsci.edu/hst/wfc3/}}
data while other \textit{HST} instruments are in use.
Observations with either the 
Cosmic Origins Spectrograph \citep[COS,][]{froning2009} or the 
Space Telescope Imager and Spectrograph \citep[STIS,][]{kimble1998}
require long integrations of a single pointing. 
During such integrations, WFC3 
\citep{kimble2008} can be used to observe targets offset by $5\farcm5$
and $4\farcm75$ from the COS and STIS primary targets, respectively. 
WFC3 has a field of view of $123'' \times 134''$ and $162'' \times 162''$
for the IR ($0\farcs13$/pixel) and UVIS ($0\farcs04$/pixel) cameras,
respectively. To date the WISP survey has observed 386 fields collectively 
covering more than 1700 sq. arcmin. 

We use both of WFC3's IR grisms:
$G_{102}$ ($0.8 - 1.1 \mu$m, $R\sim210$) and $G_{141}$ 
($1.07 - 1.7 \mu$m, $R\sim130$). To aid in extracting the spectra from the 
slitless grism images, the WISP fields were also observed in direct imaging
mode with filters chosen to match the grism spectral coverage:
F110W for $G_{102}$ and either F140W or F160W for $G_{141}$.
The WISP observing strategy depends on the length of each parallel opportunity
and can therefore vary somewhat field-to-field. In general, 
grism integration times are $\sim6\times$ those for the direct images
reflecting the sensitivities in both instrument modes.
Of the $386$ WISP fields, $117$ are also observed with the WFC3 UVIS 
camera and a subset
of the filters F475X, F600LP, F606W, and F814W 
(see Ross et al. 2016, submitted, for details).

All data are reduced with the WFC3 pipeline CALWF3 in combination 
with custom scripts that account for the un-dithered, pure-parallel
observations (\citealt{atek2010}; Ross et al. 2016, submitted). 
The UVIS images are also corrected for the charge transfer efficiency 
degradation of the WFC3/UVIS detector and
processed with customized darks based on the 
methodology of \cite{rafelski2015}. 
We use \textit{Source Extractor} \citep[version 2.5, ][]{bertin1996} for 
object detection in the direct images and 
the \verb|aXe| software
package \citep{kummel2009} to extract and calibrate the spectra.

The WISP Survey includes a shallow survey of fields observed for one to 
three continuous orbits and a deep survey of fields observed for four or more
continuous orbits. 
For these deep fields, the integration times in the two grisms are $\sim5:2$
($G_{102}:G_{141}$) in order to achieve approximately uniform sensitivity
for a line of a given flux.
In this paper we consider a subset of the deep survey, consisting 
of $48$ fields ($\sim160$ sq. arcmin\footnote{The effective grism area of the
WISP survey is $\sim3.3$ sq. arcmin, rather than the
$4.3$ sq. arcmin of the WFC3 IR channel. Area on the left side of each field
is lost because sources are not covered in the direct images -- necessary
for wavelength calibration -- while area on the right is lost 
because contaminating zero order images
cannot be identified for the spectra.}) 
reaching $5\sigma$ depths in the IR
of $m_{\mathrm{AB}} \sim26-26.8$ and UVIS $1\sigma$ depths of 
$m_{\mathrm{AB}}\geq27$.
All these fields were observed with F110W and F160W, hereafter
$\jband$ and $\hband$, respectively.
Of these $48$ fields, $21$ were observed with both F606W and F814W 
($\vband$ and $\iband$), and the remaining $27$ have only $\iband$ imaging.
The exposure times and flux limits for the 48 fields are listed
in Table \ref{tab:fields}. 
In the following sections, we describe the creation of the photometric and 
emission line catalogs used in this analysis.

\begin{table*}
\begin{center}
\begin{threeparttable}
\caption{WISP Fields used in this work, exposure times and depths}
\label{tab:fields}
\begin{tabular}{@{}lccccccccccccc}
\toprule
& Field & $t_{G102}$ & G102\footnotemark & $t_{V606}$ & $\vband$\footnotemark & $t_{I814}$ & $\iband$ & $t_{J110}$ & $\jband$ & $t_{H160}$ & $\hband$ \\
&   & [sec] & (1$\sigma$) [erg s$^{-1}$ cm$^{-2}$] & [sec] & ($1\sigma$) & [sec] & ($1\sigma$) & [sec] & ($5\sigma$) & [sec] & ($5\sigma$) \\
\midrule
& 96 & 28079 & 5.69$\times10^{-18}$ & 1500 & 28.40 & 1500 & 27.65 & 4294 & 27.39 & 1765 & 26.49 \\
& 256 & 4218 & 6.08$\times10^{-17}$ & - & - & 1500 & 27.14 & 1015 & 26.20 & 406 & 25.34 \\
& 257 & 8229 & 3.30$\times10^{-17}$ & 900 & 28.11 & 1500 & 27.67 & 1818 & 26.86 & 759 & 25.84 \\
& 258 & 6021 & 1.45$\times10^{-17}$ & - & - & 1500 & 27.73 & 1215 & 26.08 & 609 & 25.09 \\
& 260 & 7526 & 8.82$\times10^{-18}$ & - & - & 1500 & 27.70 & 1365 & 26.66 & 609 & 25.65 \\
& 261 & 6021 & 1.16$\times10^{-17}$ & - & - & 1500 & 27.55 & 1215 & 26.28 & 609 & 25.57 \\
& 271 & 4015 & 1.92$\times10^{-17}$ & - & - & 1500 & 27.36 & 809 & 26.31 & 406 & 25.31 \\
& 288 & 4218 & 1.44$\times10^{-17}$ & - & - & 1500 & 27.62 & 1015 & 26.48 & 406 & 25.43 \\
& 294 & 4218 & 1.36$\times10^{-17}$ & - & - & 1500 & 27.62 & 1015 & 26.52 & 406 & 25.44 \\
& 295 & 4318 & 1.29$\times10^{-17}$ & - & - & 1500 & 27.57 & 1015 & 26.45 & 406 & 25.41 \\
& 296 & 13544 & 8.82$\times10^{-18}$ & 1500 & 27.85 & 2500 & 27.52 & 2376 & 26.60 & 1015 & 25.90 \\
& 297 & 4518 & 1.20$\times10^{-17}$ & - & - & 1500 & 27.53 & 1015 & 26.24 & 406 & 25.31 \\
& 298 & 11835 & 1.00$\times10^{-17}$ & 1200 & 28.06 & 2000 & 27.55 & 2407 & 26.16 & 1015 & 25.60 \\
& 300 & 6921 & 1.16$\times10^{-17}$ & - & - & 1500 & 27.54 & 1262 & 26.11 & 609 & 25.61 \\
& 302 & 7023 & 1.40$\times10^{-17}$ & 900 & 27.86 & 1500 & 27.49 & 1412 & 26.69 & 812 & 25.73 \\
& 303 & 7623 & 8.09$\times10^{-18}$ & 900 & 28.16 & 1500 & 27.35 & 1315 & 26.28 & 609 & 25.62 \\
& 304 & 13038 & 4.22$\times10^{-17}$ & 1500 & 28.32 & 2500 & 27.94 & 2960 & 26.95 & 1015 & 25.95 \\
& 307 & 7323 & 1.08$\times10^{-17}$ & 900 & 27.81 & 1500 & 27.52 & 1215 & 26.61 & 812 & 25.72 \\
& 308 & 4518 & 1.85$\times10^{-17}$ & - & - & 1500 & 27.67 & 1015 & 26.53 & 406 & 25.48 \\
& 309 & 5315 & 1.29$\times10^{-17}$ & - & - & 1500 & 27.53 & 1015 & 26.52 & 456 & 25.60 \\
& 311 & 7821 & 8.82$\times10^{-18}$ & - & - & 1500 & 27.74 & 1468 & 26.42 & 609 & 25.46 \\
& 312 & 5818 & 1.10$\times10^{-17}$ & - & - & 1500 & 27.56 & 1165 & 26.81 & 609 & 25.76 \\
& 313 & 7226 & 1.49$\times10^{-17}$ & - & - & 1500 & 27.55 & 1315 & 26.82 & 609 & 25.68 \\
& 314 & 16147 & 2.84$\times10^{-17}$ & 1500 & 28.28 & 2500 & 27.81 & 2901 & 26.40 & 1543 & 26.07 \\
& 315 & 6318 & 1.35$\times10^{-17}$ & - & - & 1500 & 27.60 & 1218 & 26.78 & 609 & 25.76 \\
& 317 & 4518 & 2.34$\times10^{-17}$ & - & - & 1500 & 27.48 & 1015 & 26.40 & 406 & 25.45 \\
& 319 & 8326 & 1.40$\times10^{-17}$ & 2200 & 28.31 & 2200 & 27.76 & 1621 & 26.61 & 609 & 25.69 \\
& 320 & 12732 & 1.24$\times10^{-17}$ & 2400 & 28.23 & 2400 & 27.62 & 2782 & 26.73 & 1065 & 25.88 \\
& 321 & 12035 & 9.65$\times10^{-18}$ & 2200 & 28.46 & 2200 & 27.80 & 2154 & 26.85 & 887 & 25.94 \\
& 324 & 12135 & 9.78$\times10^{-18}$ & 2200 & 28.14 & 2200 & 27.54 & 2204 & 26.41 & 912 & 25.69 \\
& 325 & 9223 & 3.03$\times10^{-17}$ & 2400 & 28.21 & 2400 & 27.51 & 2023 & 26.78 & 762 & 25.78 \\
& 326 & 11535 & 1.27$\times10^{-17}$ & 1500 & 28.09 & 1800 & 27.55 & 2126 & 26.65 & 859 & 25.94 \\
& 333 & 15941 & 9.26$\times10^{-18}$ & 2200 & 27.24 & 2200 & 27.64 & 2866 & 26.79 & 1193 & 26.05 \\
& 340 & 7421 & 8.30$\times10^{-18}$ & 2400 & 28.49 & 2400 & 27.81 & 1721 & 26.55 & 709 & 25.88 \\
& 341 & 5921 & 1.17$\times10^{-17}$ & - & - & 1500 & 27.63 & 1215 & 26.01 & 609 & 25.55 \\
& 345 & 8021 & 9.03$\times10^{-18}$ & 2400 & 28.66 & 2400 & 27.78 & 1421 & 26.65 & 609 & 25.72 \\
& 347 & 6015 & 1.51$\times10^{-17}$ & 2400 & 27.99 & 2400 & 27.67 & 1771 & 26.39 & 1012 & 25.83 \\
& 348 & 6615 & 1.13$\times10^{-17}$ & - & - & 2000 & 27.74 & 1165 & 26.55 & 609 & 25.63 \\
& 349 & 5412 & 9.56$\times10^{-18}$ & - & - & 2400 & 27.76 & 962 & 26.33 & 609 & 25.68 \\
& 352 & 6721 & 1.19$\times10^{-17}$ & - & - & 2000 & 27.49 & 1671 & 26.64 & 684 & 25.74 \\
& 357 & 9223 & 1.11$\times10^{-17}$ & 2400 & 28.44 & 2400 & 27.80 & 2023 & 26.48 & 812 & 25.75 \\
& 360 & 7721 & 1.08$\times10^{-17}$ & - & - & 2400 & 27.76 & 1721 & 26.40 & 759 & 25.79 \\
& 364 & 6921 & 1.05$\times10^{-17}$ & - & - & 2200 & 27.85 & 1318 & 26.77 & 609 & 25.68 \\
& 368 & 6518 & 1.38$\times10^{-17}$ & - & - & 2400 & 27.78 & 1696 & 26.80 & 734 & 25.92 \\
& 369 & 9629 & 1.10$\times10^{-17}$ & 2200 & 28.51 & 2200 & 27.77 & 2276 & 26.73 & 759 & 25.86 \\
& 371 & 6618 & 1.18$\times10^{-17}$ & - & - & 2600 & 27.82 & 1218 & 26.61 & 534 & 25.47 \\
& 379 & 2809 & 1.13$\times10^{-17}$ & - & - & 2400 & 27.97 & 1721 & 26.69 & 759 & 25.94 \\
& 385 & 2406 & 1.22$\times10^{-17}$ & - & - & 1996 & 27.55 & 1165 & 26.57 & 609 & 25.68 \\
\bottomrule
\end{tabular}
\begin{tablenotes}
  \small
  \item $^{\textrm{a}}$ $G_{102}$ flux limits depend on wavelength with the 
    sensitivity of the grism. Values presented here are 
    at $\lambda=1\mu \mathrm{m}$. 
  \item $^{\textrm{b}}$ Limits are $1\sigma$ in UVIS and $5\sigma$ in the IR, 
    matching the selection criteria described in Section \ref{sec:sample}.
\end{tablenotes}
\end{threeparttable}
\end{center}
\end{table*}

\subsection{Photometric catalog}
Using AstroDrizzle \citep{drizzlepac}, we drizzle the IR direct images
onto the UVIS pixelscale ($0\farcs04$). 
We choose $0\farcs04$ because drizzling all images onto
the lower-resolution IR pixelscale ($0\farcs13$) reduces the depth in the 
UVIS images, especially for small compact objects. 
This choice only affects the final drizzled images used for photometry. 
All other data reduction and cosmic ray rejection is done on the 
original pixelscales. 

Source detection and photometry are performed with \se\ in dual image mode
using a combined detection image made from all \textit{HST} images of 
the field.
In detail,
we first convolve the UVIS images with Gaussian kernels
to match the resolution of the $\hband$
images. We then construct the detection image as the weighted 
average of the images in all four filters:
\begin{equation}
\frac{1}{n}\sum_i{\frac{I_i}{\sqrt{w_i}}}.
\end{equation}
Combining the $\jband$ and $\hband$ images produces a deeper image than
either individual filter and enables the detection of additional 
objects that would be missed by using a single filter for detection.
The UVIS images are included in our detection image to ensure that there 
is adequate coverage of all pixels. 
As we cannot dither the telescope, we do not have sufficient exposures to 
cover all IR pixels at the $0\farcs04$ pixel scale. As a result, the IR 
images drizzled to the UVIS pixel scale contain `bad' pixels that lack the 
information to have a reliable flux. These pixels have weights of zero,
and we have confirmed that they do not affect the photometry. 

We run \se\ twice for each UVIS filter: once on the original, unconvolved
image, and once on the image convolved to match the resolution of
$\hband$. The unconvolved UVIS images (at the original resolution)
are deeper than those that have been convolved. The photometry calculated
on the unconvolved images is therefore used to determine which sources
are undetected in the UVIS filters. 
The photometry from the convolved images is used in calculating
galaxy colors. 
We use circular aperture magnitudes with $r=0\farcs3 = 7.5$ pixels to 
measure galaxy colors. All other analysis is done with 
\textit{Source Extractor}'s \texttt{AUTO} elliptical apertures.

\subsection{Emission line catalog}\label{sec:emlines}
Emission line candidates are identified using an automatic algorithm that
selects groups of contiguous pixels above the continuum. 
Every candidate is then carefully inspected by two reviewers. 
This visual inspection is necessary to reject cosmic rays, hot pixels, and 
other artifacts and sources of contamination that could not be removed
during image reduction due to the lack of dithering. 
In total, 2180 emission-line galaxies have been
identified in these $48$ WISP fields.
See \cite{colbert2013} and Ross et al. (2016, submitted) 
for a detailed description of the WISP line-finding process.

While the WISP Survey is optimized to detect \ha\ and \oiii-emitters 
at $0.3 < z < 2.4$, it is also well-suited for the search for LAEs at $z>6$.
In the WISP emission line catalog, all single emission lines with no visible
asymmetry are assumed to be \ha. However, there is the possibility that some
of the faintest single-line emitters are actually LAEs at $z\geq6$. 
Most of the exposure time in the longer WISP parallel opportunities is 
devoted to $G_{102}$ observations, enabling the grism images to reach 
the sensitivities necessary for detecting $z>6$ emission lines. 
The observed EW limit of the WISP Survey \citep[40\AA][]{colbert2013}
corresponds to a low limit of $EW_0 = 5.3$\AA\ in the rest frame of a galaxy 
at $z=6.5$.
In Section \ref{sec:sample}, we describe how we select LAE candidates from 
the single-line emitters in the WISP emission line catalog.

\section{Sample Selection}\label{sec:sample}
\subsection{Choosing the selection criteria}\label{sec:criteria}
The selection criteria, which we present in Section \ref{sec:selection},
aim to select single-line emitters with colors indicative of 
a Lyman break. We choose the criteria after minimizing the number of lower-$z$
contaminanats that enter the LAE sample. 
To do so, we create a large library of simulated spectral templates based on 
the models of \cite{bc03}. From this library we generate a catalog of 
colors in the WISP bands for galaxies over $0.1\leq z \leq8.5$.
The synthetic catalog is described in Appendix \ref{sec:library}.
We then use the synthetic catalog 
to determine the effect of varying three parameters: 
(1) the range of observed emission line wavelengths we consider,
(2) the level of flux allowed in $\iband$, and
(3) the location of the selection window in color space.
In the following sections, we discuss each of these three parameters and 
their effect on the sample contamination. We define 
$f_{\mathrm{low-}z}$ as the number of lower-$z$ galaxies selected by our 
criteria divided by the total number of selected galaxies:
$f_{\mathrm{low-}z} = N_{\mathrm{low-}z} / (N_{\mathrm{high-}z} + N_{\mathrm{low-}z})$,
where $N_{\mathrm{high-z}}$ is the number of galaxies at 
$6.00 \leq z \leq 7.63$.
Only lower-$z$ galaxies in very specific redshift ranges can enter the sample
and contribute to $N_{\mathrm{low-}z}$ (see Section \ref{sec:contam} for 
details). The contamination fraction we expect for the WISP sample, denoted
$f_{\mathrm{contam}}$, is much lower than $f_{\mathrm{low-}z}$ 
because we can use the broad wavelength coverage and typical emission 
line ratios from the WISP catalog to exclude lower-redshift 
interlopers. We discuss $f_{\mathrm{contam}}$ in Section \ref{sec:contam}.
Likewise, the recovery fraction, $f_{\mathrm{recover}}$, is defined 
as the number 
of high-$z$ galaxies selected by our criteria divided by the 
total number of high-$z$ galaxies in the
full synthetic catalog. 
The maximum $f_{\mathrm{recover}}$ achieved over the full
parameter space is $0.9$.  
We choose values for the three parameters that minimize $f_{\mathrm{low-}z}$
while keeping $f_{\mathrm{recover}}$ within $10$\% of this maximum.

\subsubsection{Wavelength of emission}
The range of observed wavelengths we consider in selecting single-line
emitters determines the range of redshifts we probe for both high-$z$ LAEs
and lower-$z$ contaminants.  
The $G_{102}$ grism covers 
$0.85 \lesssim \lambda_{obs} \lesssim 1.13\mu$m, corresponding to 
$6.0 \leq z_{\mathrm{Ly}\alpha} \leq 8.3$. 
However, including sources with emission lines across this full wavelength 
range results in an unacceptable number of contaminants.
Additionally, we do not have the grism depth to detect galaxies at 
$z\gtrsim8$ (a \lya\ luminosity of 
log$_{10}(L_{\mathrm{Ly}\alpha}) = 10^{43}$ erg s$^{-1}$ at $z=8$
has a flux of $\sim1.3\times10^{-17}$ erg s$^{-1}$ cm$^{-2}$, below the 
$3\sigma$ sensitivities we reach in $G_{102}$).
We therefore apply a wavelength cut in our selection of single-line emitters.
We vary this maximum wavelength from $1.0 - 1.13\mu$m, and find that 
$\lambda_{\mathrm{obs,max}} = 1.05\mu$m results in 
$f_{\mathrm{recover}} = 0.84$ while $f_{\mathrm{low-}z}$ is just below $0.3$.

\subsubsection{Flux in $\iband$}\label{sec:iband}
A $z>6$ galaxy may emit enough flux in $\iband$ to register a signal
at greater than $1\sigma$.
For example, the Lyman break in the spectrum of a $z\sim6.6$ galaxy occurs
at $\lambda_{obs} \sim 9000$\AA, $\sim1000$\AA\ blueward of the edge of
the $\iband$ filter. 
With imaging in at most one additional filter blueward of $\iband$,
we cannot confirm that a $z\geq6$ galaxy with some flux in $\iband$ 
has truly dropped out of the optical. 
The level of $\iband$ source flux 
above the background we allow in selecting candidates will therefore 
strongly affect the expected contamination fraction.
We consider $\iband$ detection thresholds from $1\sigma - 3\sigma$, and 
find that $1\sigma$ is the best option in terms of recovery and 
contamination fractions ($f_{\mathrm{recover}} = 0.83$,
and $f_{\mathrm{low-}z} = 0.29$). Including galaxies that are 
nondetections at the $1.5\sigma$ level increases the contamination 
to $f_{\mathrm{low-}z} \sim 0.4$. There are two single-line emitters in 
the WISP catalog that are nondetections in $\iband$ at $1.5\sigma$.
We conservatively reject these to avoid contamination 
from faint, lower-redshift interlopers. 
We include their spectra and 
direct image stamps in Appendix \ref{app:others}.

\subsubsection{Selection window}
\begin{figure}
\begin{center}
\includegraphics[width=0.5\textwidth]{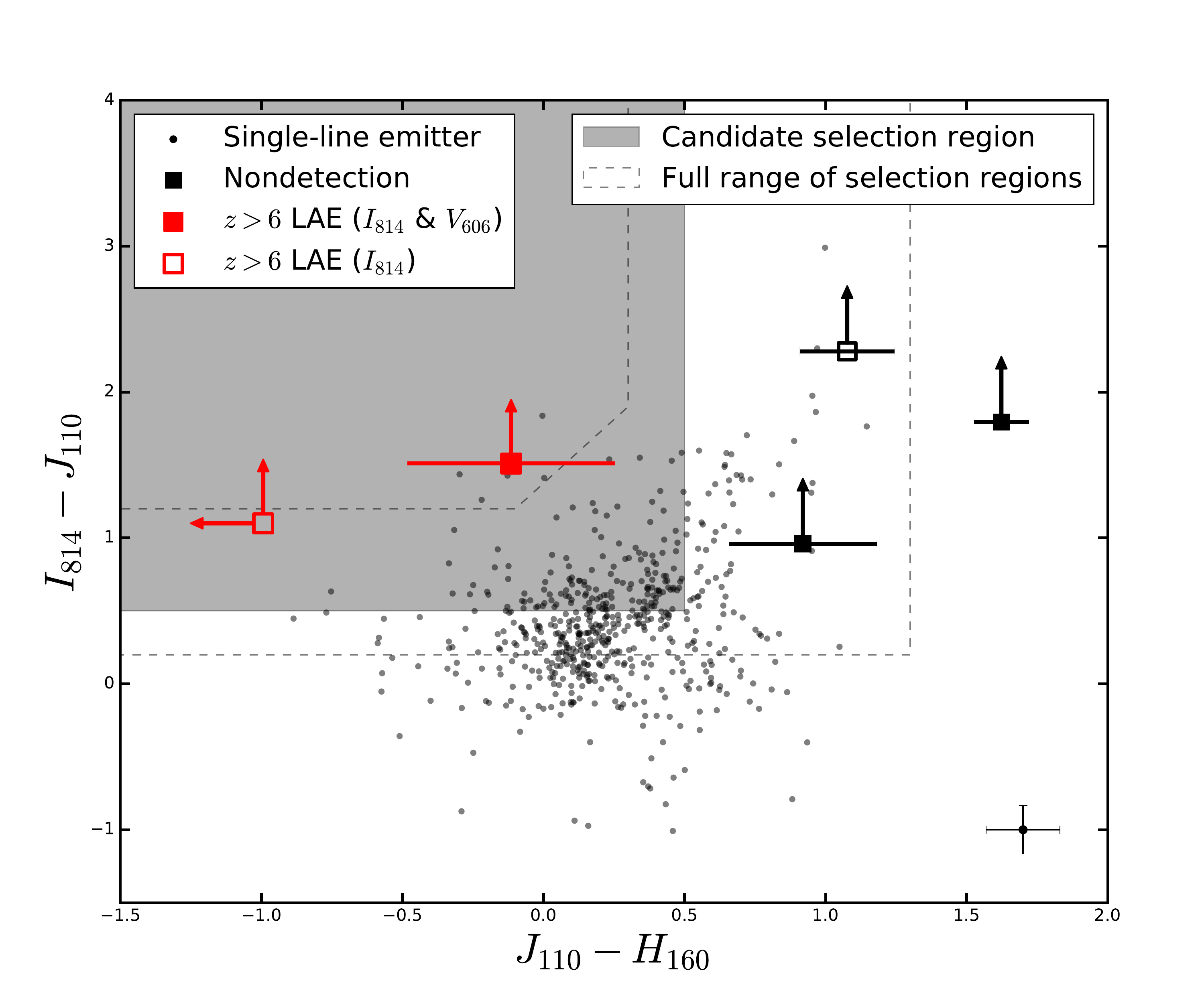}
\caption{$\iband - \jband$ vs. $\jband - \hband$ color-color plot 
used to identify $z>6$ dropout galaxies. The grey region shows
the candidate selection region. The dashed lines indicate 
the range of selection regions we consider in order to minimize 
$f_{\mathrm{low-}z}$. All galaxies plotted here are single-line 
emitters with $\lambda_{obs}\leq1.05\mu$m. 
Galaxies that are undetected in the UVIS filter(s) are plotted as squares:
either solid if both $\vband$ and $\iband$ imaging are available or 
open if we have only $\iband$. 
The selected LAE candidates
are indicated by red squares.
$1\sigma$ error bars are shown, and nondetections are plotted at the
$1\sigma$ magnitude limits calculated for their fields. Typical error bars
for the UVIS-detected galaxies are plotted in the lower right corner.
There are three additional UVIS nondetections (black squares) that 
lie outside of the selection region. 
We discuss these sources in Appendix \ref{app:others} but do not
consider them $z>6$ LAE candidates due to their red IR colors. 
\label{fig:ccd}
}
\end{center}
\end{figure}

Figure \ref{fig:ccd} shows the color-color plot used to identify 
$z>6$ dropout galaxies. In choosing the selection criteria, 
we vary the shape and size of the color selection window, 
covering the range indicated by the dashed lines in Figure \ref{fig:ccd}.
The red limit in ($\jband - \hband$) has the largest effect on both 
$f_{\mathrm{recover}}$ and $f_{\mathrm{low-}z}$. We consider windows as red
as $(\jband-\hband) = 1.3$ and 
choose $0.5$ with $f_{\mathrm{recover}} = 0.8$
and $f_{\mathrm{low-}z} = 0.22$.
Only 1\% of the $z<6$, UVIS-undetected galaxies in the selection window 
come from the lower-right corner of our selection region. For simplicity, 
we therefore use a rectangular selection region rather than 
the more complicated windows used by, e.g. \cite{oesch2010} and 
\cite{bouwens2015}. 
The requirement for emission line detection in our sample 
removes any concern over contamination from this region.

\subsection{Selection criteria}\label{sec:selection}
A WISP single-line emitter is identified as a $z>6$ LAE candidate if and 
only if it:
\begin{enumerate}
  \item has a single emission line at $\lambda_{obs} \leq 1.05\mu$m;
  \item is detected with a signal-to-noise (S/N) ratio in $\jband$ of
    $(S/N)_{\jband} \geq 5$;
  \item has colors consistent with those of a $z\geq6$ galaxy:
\begin{align*}
(\iband - \jband) &\geq 0.5  \\
(\jband-\hband) &\leq 0.5;
\numberthis \label{eqn:color} \\
\end{align*}
  \item is undetected at the $1\sigma$ level in all available UVIS filters 
    (``UVIS dropouts'').
\end{enumerate}
We derive the color selection criteria
in Appendix \ref{sec:library}
and identify it by the shaded region in Figure \ref{fig:ccd}. 
These color criteria are comparable to those used in similar searches for 
$z\sim6-7$ galaxies, e.g. \cite{oesch2010}, \cite{schenker2013} and 
\cite{bouwens2015} who employ color cuts of $(I-J)>0.7-1$ and 
$(J-H) < 0.4- 1$ in the same or similar filters as those used here.

We adopt the $1\sigma$ limiting magnitudes for all galaxies that are
$1\sigma$ nondetections\footnote{In order to confirm the 
detection or non-detection of each source, 
we measure the residual flux of the surrounding sky.
We place the same size circular apertures ($r=0\farcs3=7.5$ pix) 
randomly on the 
sky around each source out to a radius of $5"$. 
A source that is fainter than $\geq84\%$ of these sky apertures 
is considered a $1\sigma$ nondetection in that filter.}
and plot them in Figure \ref{fig:ccd} as limits.
Each LAE candidate is then inspected by eye to confirm the emission lines are
real and the galaxy is truly undetected in the UVIS filters.

\begin{figure}
\begin{center}
\includegraphics[width=0.5\textwidth]{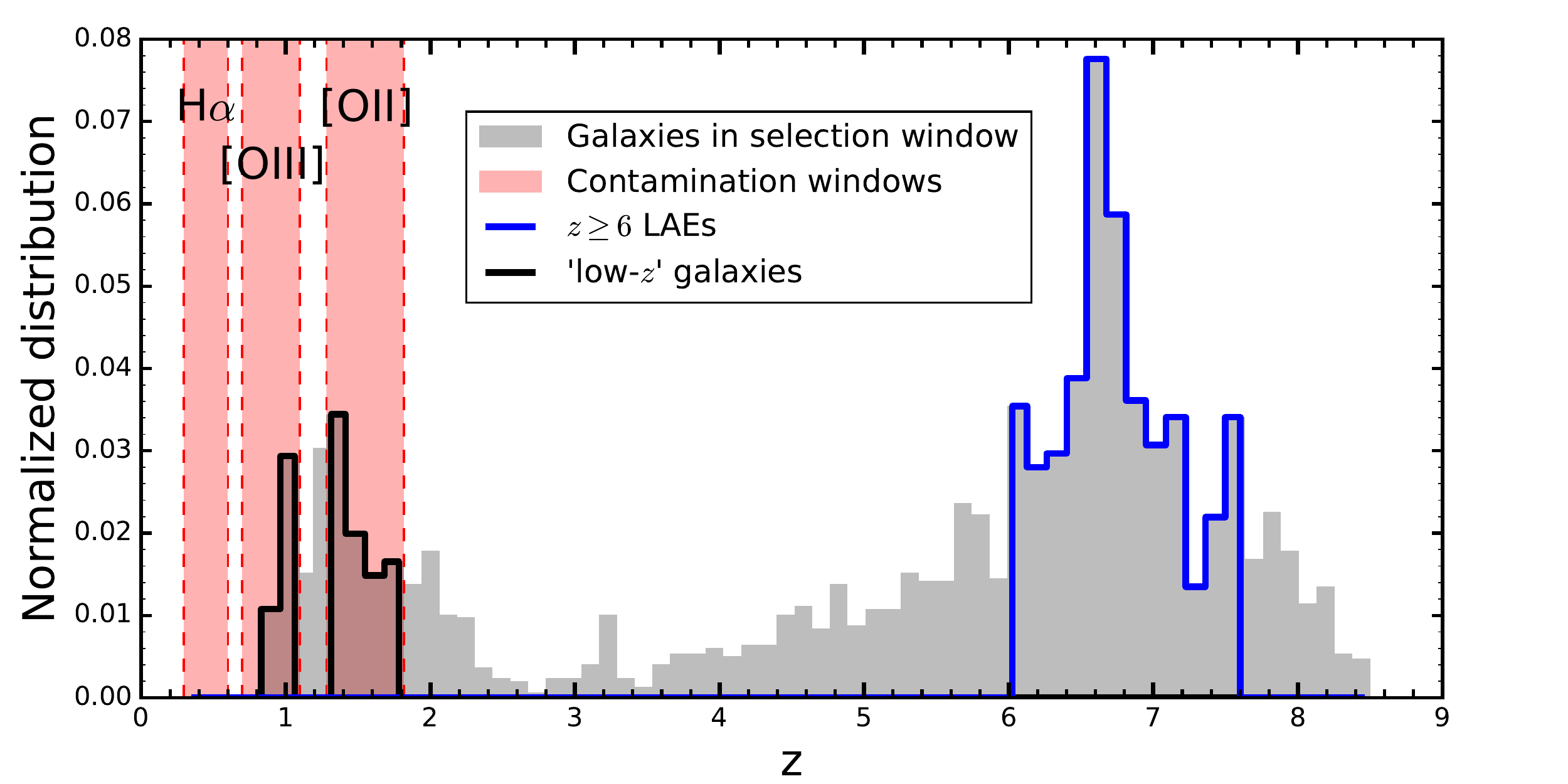}
\caption{
Characterization of the expected contamination in the LAE 
sample from the synthetic catalog derived in Appendix \ref{sec:library}.
The redshift distribution of 
UVIS nondetections in the color selection window is shown in grey.
The distribution of the 
high-$z$ population we are targeting ($6.0 \lesssim z \lesssim 7.6$) is
outlined in blue. The shaded red bands indicate redshift windows capable of 
contributing contaminants to the high-$z$ sample, and the distributions of 
galaxies within these bands are outlined in black. 
These galaxies account for 29\% ($f_{\mathrm{low-}z}$) of the total 
selected sample ($blue + black$).
WISP has the spectral coverage to detect multiple emission lines in the 
spectra of these ``low-$z$'' galaxies. As a result, only  
those with additional emission lines fainter than the WISP
sensitivity would contaminate the LAE sample ($\sim2\%$, 
see Section \ref{sec:contam}). 
\label{fig:zdist}
}
\end{center}
\end{figure}

\section{Contamination}\label{sec:contam}
High-redshift samples selected via the Lyman break technique can be
highly contaminated by lower-redshift interlopers. 
If, for example, there is any overlap between the red and blue filters,
as between $\iband$ and $\jband$, the dropout color as a function of
redshift can flatten into a `color plateau' \citep{stanway2008}.
Photometric uncertainty can
scatter lower-redshift galaxies into the
selection window while simultaneously scattering real high-redshift sources
out. 
Additionally, Balmer break galaxies at $1.3\lesssim z \lesssim2.5$ 
can mimic the colors of $z\geq6$ 
galaxies in filters chosen to bracket the Lyman break
\citep[e.g.,][]{mobasher2005,henry2009,pirzkal2013}.

We characterize the expected contaminants to the LAE sample in Figure
\ref{fig:zdist}.
The grey histogram 
shows the redshift distribution of 
sources from our synthetic catalog that lie in the selection window 
and are nondetections in both $\vband$ and $\iband$. These sources
range from $0.7\lesssim z \lesssim 8.5$ and include a large number of 
$z < 6$ galaxies.

However, as described in Section \ref{sec:criteria}, we consider single-line
emitters with lines in the range $0.85 \lesssim \lambda_{obs} \leq 1.05\mu$m.
We remind the reader that the red wavelength limit is chosen to minimize 
contamination and the blue limit is determined by the rapid drop in the grism 
sensitivity.
Given these wavelengths, we are probing \lya\ at 
\begin{equation}
6.00 \lesssim\ z_{\mathrm{Ly}\alpha}\ \lesssim\ 7.63.
\label{eqn:lya}
\end{equation}
Galaxies in this redshift range are identified by the blue histograms in 
Figure \ref{fig:zdist}.
The wavelength selection also restricts contaminants to very narrow 
redshift ranges (identified as shaded red regions in Figure \ref{fig:zdist}):
\begin{itemize}
\item \ha\ between $0.3 \lesssim\ z_{\mathrm{H}\alpha}\ \lesssim 0.6$,
\item \oiii\ between $0.7 \lesssim\ z_{\mathrm{[O~III]}}\ \lesssim 1.1$, and
\item \oii\ between $1.3 \lesssim\ z_{\mathrm{[O~II]}}\ \lesssim 1.8$.
\end{itemize}
The galaxies from these redshift regions are outlined in black,
and the resulting fraction of lower-$z$ sources that are selected alongside
the $z>6$ galaxies is $f_{\mathrm{low-}z} = 0.29$.
This fraction, however, is much higher than the contamination fraction 
we expect for the LAE sample, $f_{\mathrm{contam}}$,
which is limited to lower-redshift galaxies with very specific emission 
line ratios (see Section \ref{sec:fem}).
In what follows, we show that very few galaxies ($\lesssim 2$\%)
in these redshift ranges could contaminate our results.

\begin{figure*}
\begin{center}
\includegraphics[width=\textwidth]{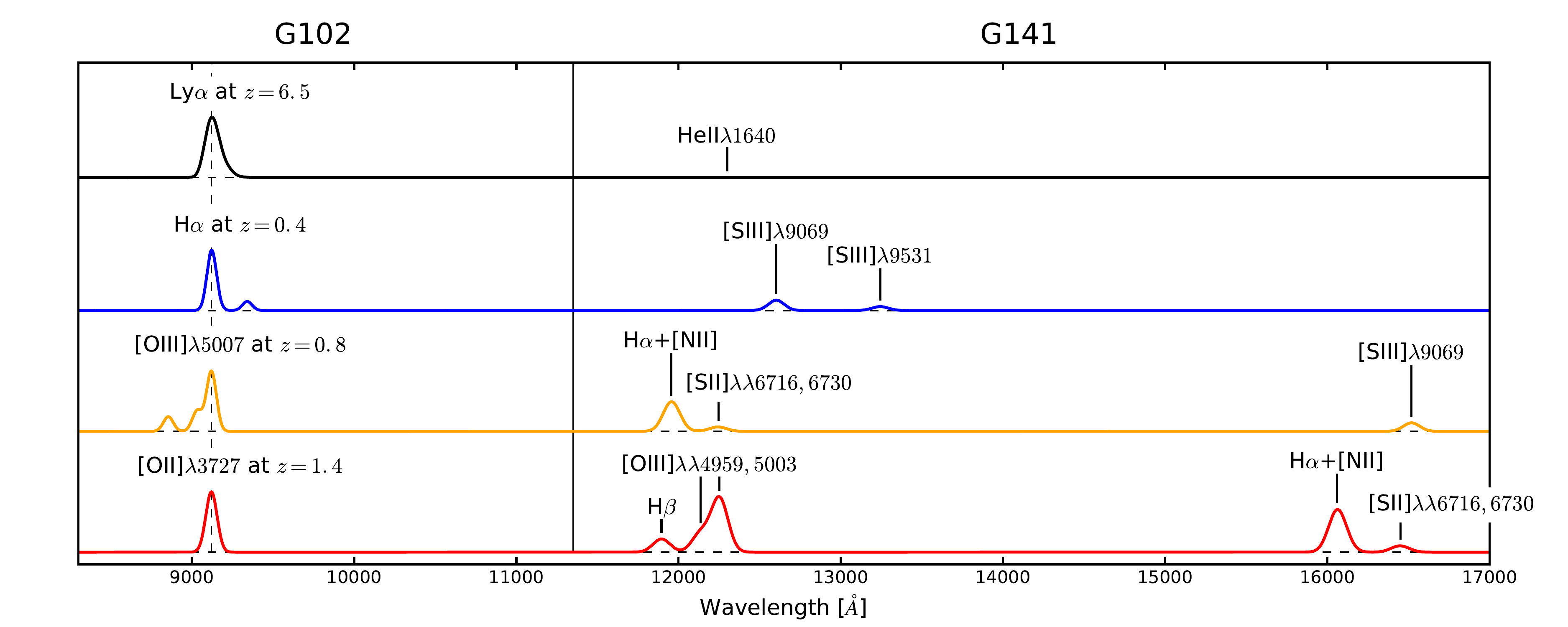}
\caption{An illustrative example of how the full wavelength coverage of WISP
can be used to rule out lower-$z$ interlopers. We show here three
possible misidentifications, where 
\ha\ at $z=0.4$ (blue), \oiii\ at $z=0.8$ (orange) and \oii\ at $z=1.4$ (red)
are mistaken for \lya\ at $z=6.5$ (black).
In each case, we expect to detect additional emission at longer wavelengths.
We can therefore rule out all lower-$z$ contaminants but those where the
redder emission lines fall below the WISP sensitivity.  
The spectra plotted here have been convolved to match
the resolutions of the $G_{102}$ and $G_{141}$ grisms. 
\label{fig:speclines}
}
\end{center}
\end{figure*}

If the LAE candidates are lower-$z$ contaminants, we would expect
to detect additional emission lines in the $G_{141}$ grism.
For the purposes of illustration, we plot a simplified emission-line spectrum
in Figure \ref{fig:speclines} at four redshifts:
$z=6.5$ (black), $1.4$ (red), $0.8$ (orange), and $0.4$ (blue).
Because we select the LAE candidates from the sample of single-line 
emitters in the WISP catalog, no additional emission lines are detected 
in the spectra of these galaxies over the full 
wavelength coverage of the WISP survey.
Therefore, if they are lower-redshift galaxies, 
the additional emission lines must 
fall below the sensitivity of the WISP 
spectra\footnote{It is also possible that a second emission line could 
fall on the 
overlapping edge of the two grisms and be lost due to the drop in 
sensitivities. Such galaxies would be mistakenly identified as single-line
emitters. This would only occur, however, for a small subset of emission
lines in the range $1.13 \lesssim \lambda_{obs} \lesssim 1.15\mu$m,
and does not affect the LAE candidates.
If the candidates are lower-$z$ galaxies, the additional emission lines 
such as \oiii\ and \ha\ would not fall in the affected wavelength range. 
}.
We discuss this possibility in the following section.

\subsection{Emission line misidentification}\label{sec:fem}
In the case that the \lya\ emission line has been misidentified, 
it would likely be either \ha, \oiii, or \oii\ from lower-$z$ galaxies.
It is difficult to rule out \ha-emitters based on the candidates' 
spectra because there are few emission lines
detected redward of \ha\ in WISP spectra. From our synthetic catalog,
however, we find that all galaxies in the redshift range 
in which \ha\ can enter our sample are detected in the UVIS images 
(see Figure \ref{fig:zdist}). 
Therefore, we conclude that contamination from \ha\ is negligible.

\begin{figure*}
\begin{center}
\includegraphics[width=\textwidth]{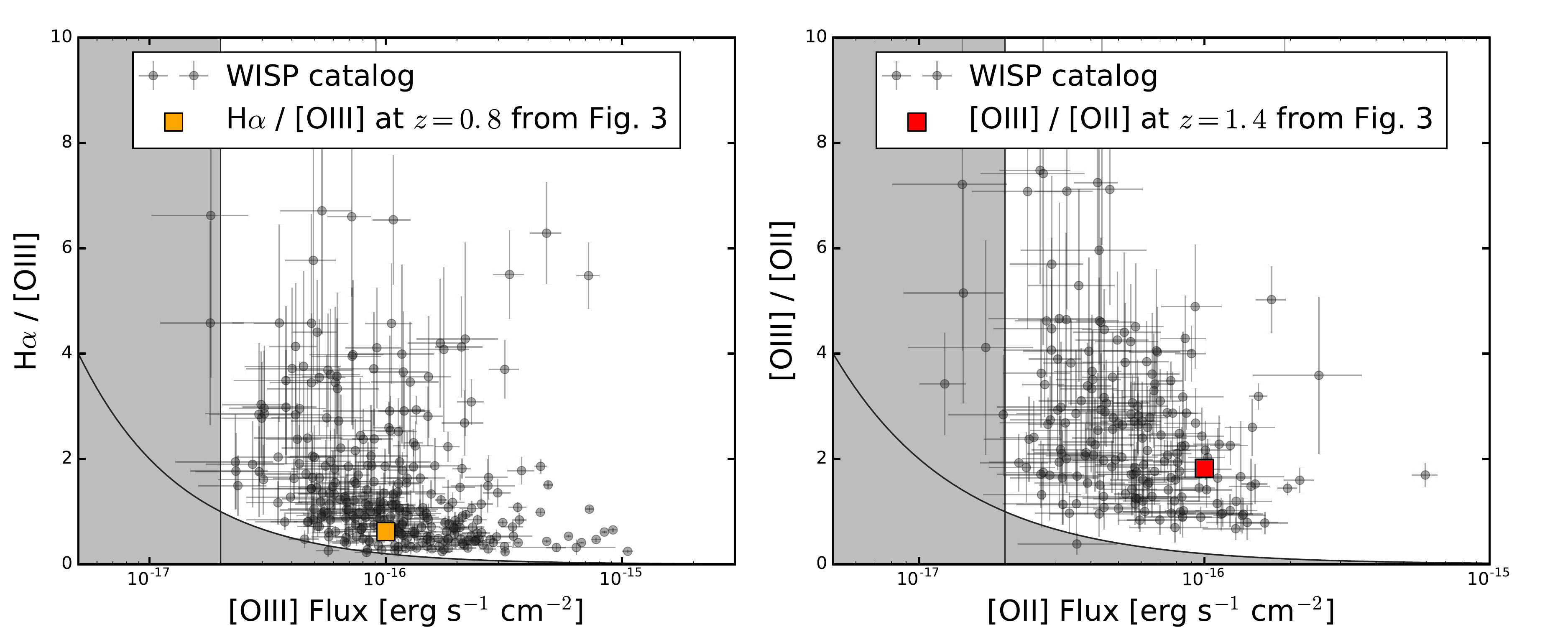}
\caption{The line ratios for all sources in the WISP catalog for which 
the bluer of the two emission lines lies in our wavelength selection 
range ($0.85 \lesssim \lambda_{obs} \leq 1.05\mu$m). 
\textit{Left:} \ha/\oiii\ as a function of \oiii\ flux. 
\textit{Right:} \oiii/\oii\ as a function of \oii\ flux.
The shaded areas show the regions where one of the emission lines
is below the detection limit of the WISP spectra, 
assuming $2\times10^{-17}$ \ecs\ as
the limiting line flux. 
For reference, the orange and red squares show the line ratios 
for the $z=0.8$ (orange) and $z=1.4$ (red) spectra in Figure \ref{fig:speclines}.
\label{fig:lineratios}
}
\end{center}
\end{figure*}

For the remaining likely contaminants, \oiii- and \oii-emitters,
we look to the full WISP emission line catalog for information on the 
expected line ratios of galaxies at $z\lesssim2$.
The \oiii\ doublet is blended at the resolution of the WFC3 grisms, and 
so we use \oiii\ to refer to \oiii$\lambda\lambda4959,5007$ in what follows. 
Similarly, \ha\ fluxes are not corrected for \nii.
Figure \ref{fig:lineratios} shows the \ha/\oiii\ (left panel) and 
\oiii/\oii\ (right) line ratios for sources in the WISP catalog. 
Here we consider all sources for which the bluer of the two emission lines
lies in our wavelength selection range 
($0.85 \lesssim \lambda_{obs} \leq 1.05 \mu$m). 
For example, if an emission line
at $\lambda_{obs}=0.9\mu$m is \oiii\ instead of \lya, then we would 
expect to detect \ha\ at $\lambda_{obs}=1.18\mu$m.
The shaded areas in Figure \ref{fig:lineratios} show the approximate 
regions of lower limits in the case where \ha\ (left) or \oiii\ (right) 
are undetected in the WISP spectra.
For both cases we adopt $2\times10^{-17}$ erg s$^{-1}$ cm$^{-2}$ as 
the limiting line flux, which is the average line flux 
limit in the deep WISP  
fields\footnote{Emission lines with fluxes below this value are either in 
the deepest fields or are detected at lower significance than the rest
of the catalog. If there are multiple
emission lines in the spectra, the secondary lines are measured even if
they are at lower S/N.}.
The line ratios for the $z=0.8$ and $z=1.4$
spectra plotted in Figure \ref{fig:speclines} are indicated in Figure 
\ref{fig:lineratios} by orange and red squares,
respectively, and are meant as illustrative examples for reference only. 

The median line flux ratios in the WISP catalog are 
\ha/\oiii$\sim1$ and \oiii/\oii$\sim2.5$, consistent with previous
results from both individual galaxies \citep{mehta2015} 
and stacked WISP spectra \citep{dominguez2013,henry2013b}.
If the LAE candidates are lower-$z$ contaminants, their line ratios
would have to be extreme: \ha/\oiii\ or \oiii/\oii\ $\lesssim0.2$ for fluxes of
$\sim1\times10^{-16}$ erg s$^{-1}$ cm$^{-2}$. In other words,
if the single detected emission line is actually \oiii\ (\oii), then 
\ha\ (\oiii) must be a factor of $\gtrsim5$ fainter to fall below our 
sensitivity. 

As can be seen from the right panel of Figure \ref{fig:lineratios},
\oii-emitters will not contaminate the LAE sample. 
At line fluxes of $f=1\times10^{-16}$ erg s$^{-1}$ cm$^{-2}$, virtually
no galaxies have \oiii/\oii $< 0.2$. Because \oiii\ is so much stronger
than \oii, we are guaranteed not to mistake \oii\ for \lya. 
On the other
hand, $2$\% of galaxies with 
$7\times10^{-17} \leq f_{\mathrm{[OIII]}} \leq 2\times10^{-16}$ \ecs\ 
have \ha/\oiii $<0.25$. 
Hence, contamination from \oiii-emitters
is very rare but still a possibility. 

There is always the possibility that hot pixels, cosmic rays, 
and other artifacts are masquerading as emission lines, although we expect
only an $\sim8.5$\% contamination rate due to false emission lines
\citep{colbert2013}. 
Due to the different dispersion solutions, the spectra 
do not fall on exactly the same pixels in $G_{102}$ and $G_{141}$. 
By comparing the full dispersed images in each grism, 
we performed a check for hot pixels, detector artifacts,
and persistence
that could have been incorrectly identified as emission lines. 
We also inspected each individual grism exposure to ensure that 
we are not detecting cosmic rays.
Finally, we have checked that the emission lines in our sample
are not zeroth orders from nearby objects. All galaxies in our sample lie
sufficiently far from the right edge of the detector, 
and so zeroth order images are easily identified for their spectra. 

Considering contamination from both lower-$z$ and false emission lines,
we expect the contamination fraction in the LAE sample to be no higher than
$f_{\mathrm{contam}} \sim 2-8$\%.

\section{Results}\label{sec:results}

\subsection{$z>6$ LAEs}
We find two robust candidates according to the selection criteria described in 
Section \ref{sec:sample}, \textit{WISP368} at $z=6.38$ and 
\textit{WISP302} at $z=6.44$.
Table \ref{tab:cand} summarizes the spectroscopic and photometric properties.
For both sources, a large fraction of the flux density 
in $\jband$ is due to the 
emission line, and we include this fraction, 
$f_{J110}^{\mathrm{neb}} / f_{J110}^{\mathrm{total}}$ in Table \ref{tab:cand}.
We calculate the rest frame EW of \lya\ using the $\hband$ magnitudes 
for the continuum measurement, as the $\hband$ magnitudes are not
contaminated by \lya\ emission. In the case of $\textit{WISP368}$,
we adopt the $3\sigma$ $\hband$ magnitude limit.
We also present the UV absolute magnitudes calculated at rest frame
wavelengths of 
$1500$\AA\ and $2000$\AA, corresponding to $\jband$ and $\hband$, respectively.
$M_{1500}$ is calculated using the $\jband$ magnitude corrected for 
the emission line flux.
Still, $M_{1500}$ is uncertain and so we do not calculate 
the UV slope $\beta$.
Finally, we measure $3\sigma$ upper limits for
 He{\small II}$\lambda1640$ and C{\small III]}$\lambda1909$ fluxes.

\begin{figure}
\begin{center}
\includegraphics[width=0.5\textwidth]{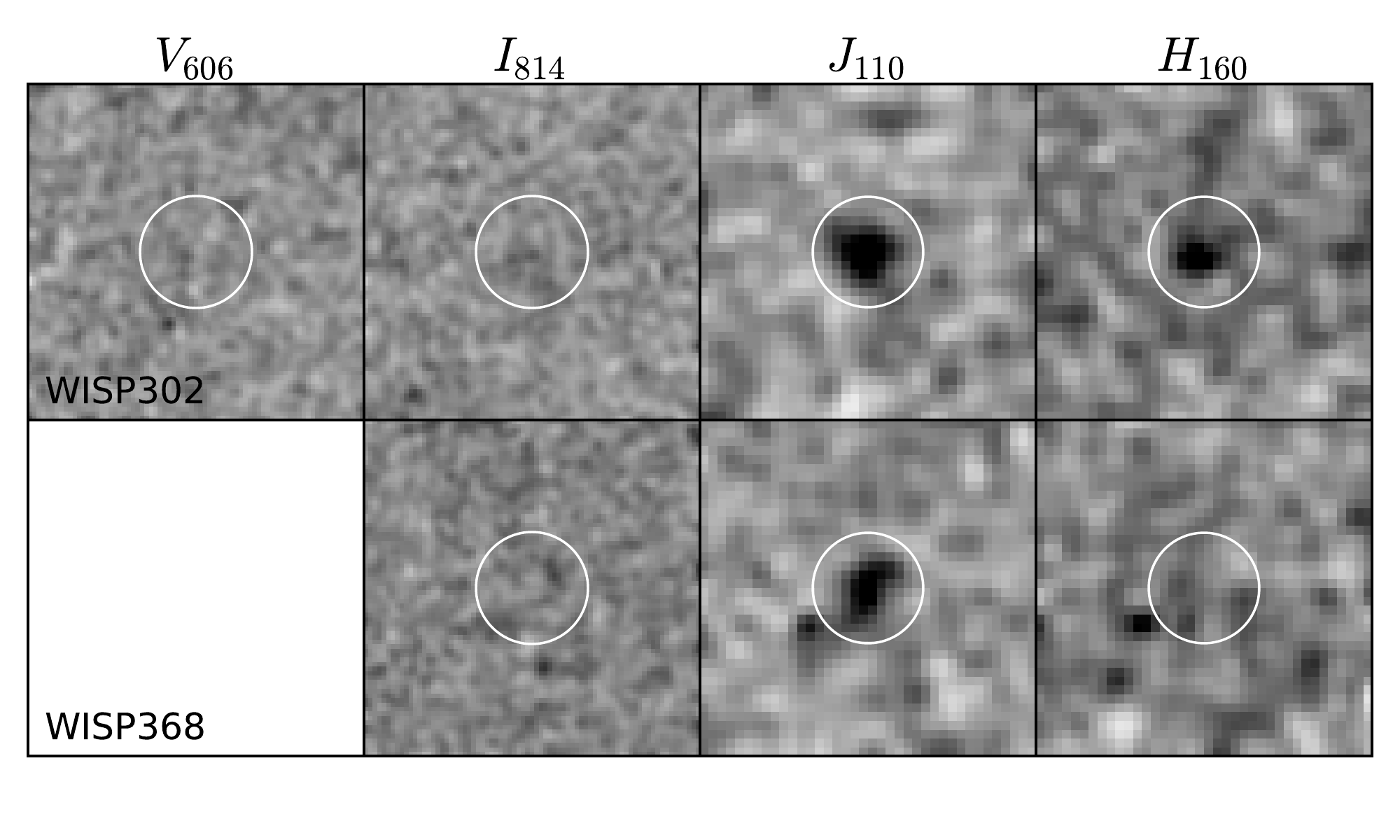}
\caption{Direct image postage stamps of the LAEs. Columns
show, from left to right, $\vband$, $\iband$, $\jband$, and $\hband$. 
All stamps are $3^{\prime\prime}$ on a side. 
The UVIS stamps have a pixel scale of 
$0\farcs04$/pix and the IR stamps have been re-drizzled here onto a
pixel scale of $0\farcs08$/pix. All stamps are smoothed with a 
Gaussian kernel with $\sigma=1$ pixel ($\sigma=0\farcs04$ and $0\farcs08$
in the UVIS and IR, respectively). The white circles are $0\farcs5$ in 
radius.
\label{fig:stamps}
}
\end{center}
\end{figure}

We present images of each candidate in Figure \ref{fig:stamps}.
The stamps are $3"$ on a side and are each smoothed with a 
Gaussian kernel with $\sigma=1$ pixel. 
The candidates have dropped out of the UVIS filters: $\vband$ and $\iband$ for
\textit{WISP302} and $\iband$ for \textit{WISP368}.

\begin{figure*}
\begin{center}
\includegraphics[width=\textwidth]{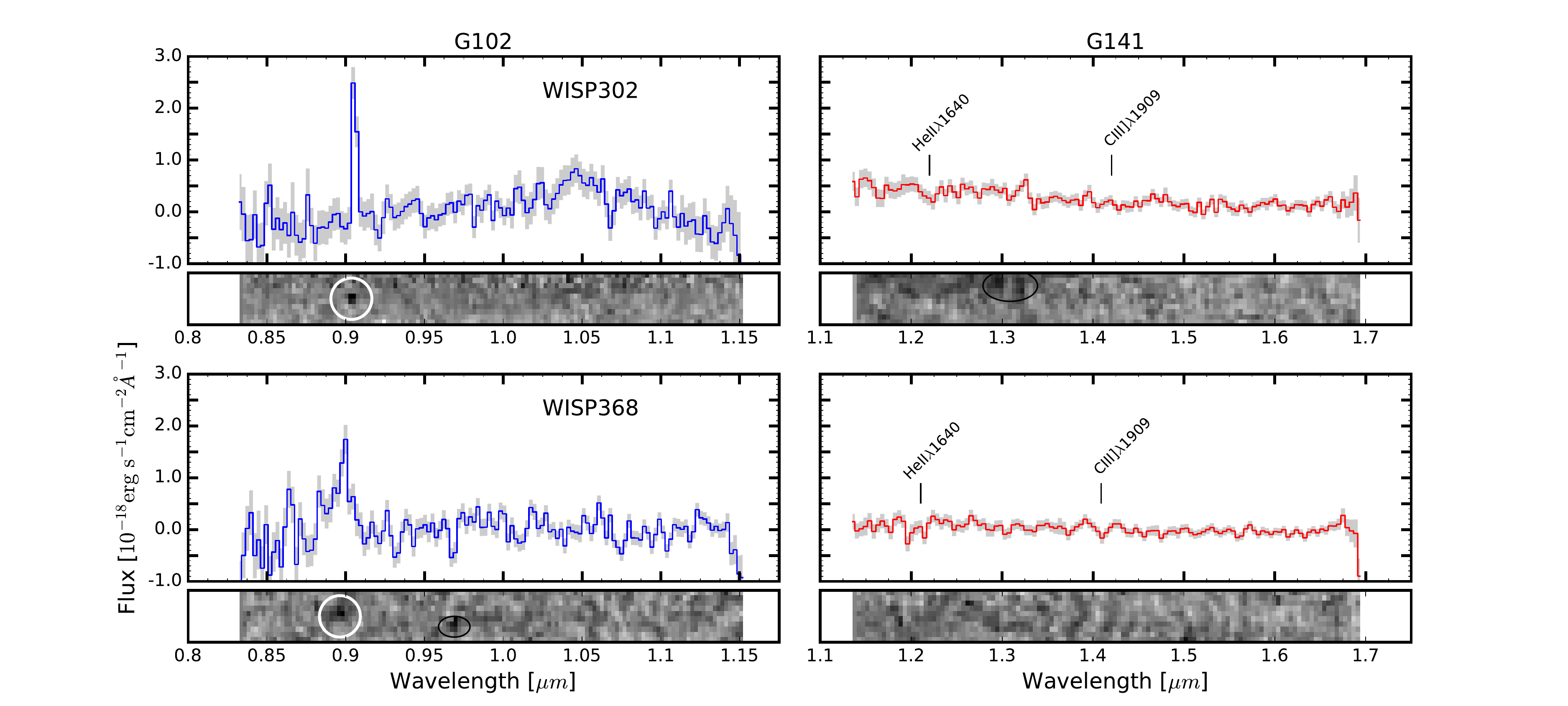}
\caption{The one- and two-dimensional spectra of the LAE candidates in 
$G_{102}$ (left column) and $G_{141}$ (right). In the one-dimensional spectra,
the $1\sigma$ errors are plotted in grey. The grism spectral stamps have 
been smoothed by $0.5$ pixels. The \lya\ emission lines are circled in 
white, and nearby ``emission features'' are identified in black.
\label{fig:spectra}
}
\end{center}
\end{figure*}

Figure \ref{fig:spectra} shows the one- and two-dimensional $G_{102}$ 
and $G_{141}$ spectra for each candidate. The \lya\ emission lines 
are circled in white.
Additional ``emission features'', identified in black in Figure 
\ref{fig:spectra}, are present in the dispersed images 
of both objects. After careful examination of the images, we conclude 
that these features belong to nearby sources.
The wavelengths of He{\small II}$\lambda1640$ and 
C{\small III]}$\lambda1909$ are also labeled.

We use a Monte Carlo process to measure the line fluxes: we fit the line 
with a Gaussian, allow the flux at each wavelength step to vary within the
uncertainties, and refit. The flux uncertainties 
in Table \ref{tab:cand} are the $1\sigma$
dispersion in the distribution of measured fluxes. 

\begin{table*}
\begin{center}
\begin{threeparttable}
\caption{LAEs}
\label{tab:cand}
\begin{tabular}{@{}lcccccccccccccc}
\toprule
& ID & R.A. & Decl. & $z_{\mathrm{Ly}\alpha}$ & $f_{\mathrm{Ly}\alpha}$ & EW$_0$ & $L_{\mathrm{Ly}\alpha}$ & $f_{\mathrm{HeII}}$\tnote{a} & $f_{\mathrm{CIII]}}$\tnote{a} \\
& & (J2000) & (J2000) & & [$10^{-17}$ erg/s/cm$^2$] & [\AA] & [$10^{43}$ erg/s] & [$10^{-17}$ erg/s/cm$^2$] & [$10^{-17}$ erg/s/cm$^2$] \\
\midrule
& WISP302 & 02:44:54.72 & $-$30:02:23.3 & 6.44 & $9.9\pm5.8$ & $798\pm531$ & 4.67  & $<6.0$ & $<4.3$ \\
& WISP368 & 23:22:32.26 & $-$34:51:03.7 & 6.38 & $10.2\pm3.9$ & $<1172$ & 4.71 & $<3.9$ & $<2.8$ \\
\toprule
\\ 

& ID & $\vband$\tnote{b} & $\iband$\tnote{b} & $\jband$\tnote{b} & $f_{J110}^{\mathrm{neb}} / f_{J110}^{\mathrm{total}}$ & $\hband$\tnote{b} & $M_{1500}$\tnote{c} & $M_{2000}$ \\
&  & [mag] & [mag] & [mag] &  & [mag] & [mag] & [mag] \\
\midrule
& WISP302 & $>27.86$ & $>27.49$ & $26.0\pm0.11$ & $0.67$ & $26.1\pm0.35$ & $-19.6$ & $-20.7$ \\
& WISP368 & - & $>27.78$ & $26.3\pm0.20$ & $0.96$ & $>27.67$ & $-17.0$ & $>-19.1$ \\
\bottomrule 
\end{tabular}
\begin{tablenotes}
  \small
  \item[a] Flux limits are $3\sigma$.
  \item[b] Total magnitudes as measured by 
    \textit{Source Extractor}'s \texttt{AUTO} elliptical apertures.
  \item[c] $M_{1500}$ is calculated using the $\jband$ magnitude corrected for 
    the emission line flux.
\end{tablenotes}
\end{threeparttable}
\end{center}
\end{table*}

\subsection{What are the redder UVIS dropouts?}\label{sec:others}
There are three additional
sources (black squares in Figure \ref{fig:ccd}) that are too red to fall in our 
selection window yet meet all other criteria. As single-line emitters that 
drop out of the UVIS filters, these sources are worth further consideration.
These three galaxies are small and compact and lie in the region of color space
that contains a higher fraction of the dusty, red, lower-$z$ galaxies in our 
synthetic library.
Such galaxies
may belong to a population of galaxies at $z\sim1.5-1.8$ that are either 
very dusty or have strong $4000$-\AA\ breaks, causing them to mimic the 
broadband colors of the Lyman break.
The emission line is most likely \oii.
According to \cite{maiolino2008}, the observed limit of 
\oiii/\oii\ (\oiii/\oii\ $\lesssim0.2-0.25$) would imply 
high metallicities at these redshifts. 
Understanding the nature of these sources  
is very important, but beyond the scope of this paper.
We do not consider these three sources in our analysis and instead present
their spectra and direct image stamps in Appendix \ref{app:others}. 

\subsection{Calculation of the number density}\label{sec:n}
Given the two LAEs presented above, we 
calculate the number density as:
\begin{equation}
n_{\mathrm{LAE}} = \underset{i}\sum \ C_i \ \frac{1}{V_i},
\end{equation}
where $C_i$ is an emission line dependent completeness correction and 
$V_i$ is the volume within which the emission line could be detected.
The dominant sources of incompleteness in the WISP survey are 
(1) confusion from nearby objects and partially- or fully-overlapping spectra,
and (2) the failure of the line-finding process to identify emission lines. 
\cite{colbert2013} simulate the full emission line identification process
and derive completeness corrections that depend on emission line equivalent
width (EW) and line flux S/N. 
We do not detect continua in the LAE spectra, and so use the completeness
corrections appropriate for the highest-EW lines in the WISP catalog.

The WISP fields reach a range of depths.
In determining $V_i$ we use a modified version of the $V_{\mathrm{MAX}}$ 
method \citep{felten1977} that depends on the $G_{102}$ sensitivity limit
reached in each WISP field. 
The maximum volume within which a galaxy with the given 
absolute \lya\ luminosity could be detected
is: 
\begin{equation}
V_{\mathrm{MAX}}\mathrm{(LAE)} = \overset{N_{fields}=48}{\underset{i=1}{\sum}} \ \Omega_i \int^{z_{U,i}}_{z_{L,i}}  \frac{dV}{dz} \mathrm{ d}z.
\label{eqn:vmax}
\end{equation}
Here, $\Omega_i$ is the effective area of one WISP 
field (3.3 sq. arcmin), 
$z_L$ is the lower redshift at which the galaxy's \lya\ luminosity 
would fall below 
the G102 grism sensitivity, which cuts off steeply at the blue end. $z_U$ 
is the minimum of (1) the redshift at which the galaxy's line flux would 
fall below the sensitivity on the red side, and (2) the upper redshift
limit set by our 
selection criteria, $z=7.63$. The \lya\ LF, however, is expected to 
evolve rapidly above $z=7$. For the purposes of comparing our number counts
with those of other studies at $z\sim6.5$, we limit this maximum redshift
to $z=7$. We present an upper limit for the volume at 
$7.0 \leq z \leq 7.63$ in Section \ref{sec:laez7}.

The integration limits in Equation (\ref{eqn:vmax}) are calculated 
for each of the $48$ fields because 
the depth is not uniform across all fields.
This corresponds to different redshift limits, $z_U$ and $z_L$, and 
therefore different volumes probed in each field.
Smaller effective volumes are probed in the shallower fields.
In some cases, the emission line may not be detectable in a shallow
field at any of the relevant redshifts. 

\begin{figure}
\begin{center}
\includegraphics[width=0.5\textwidth]{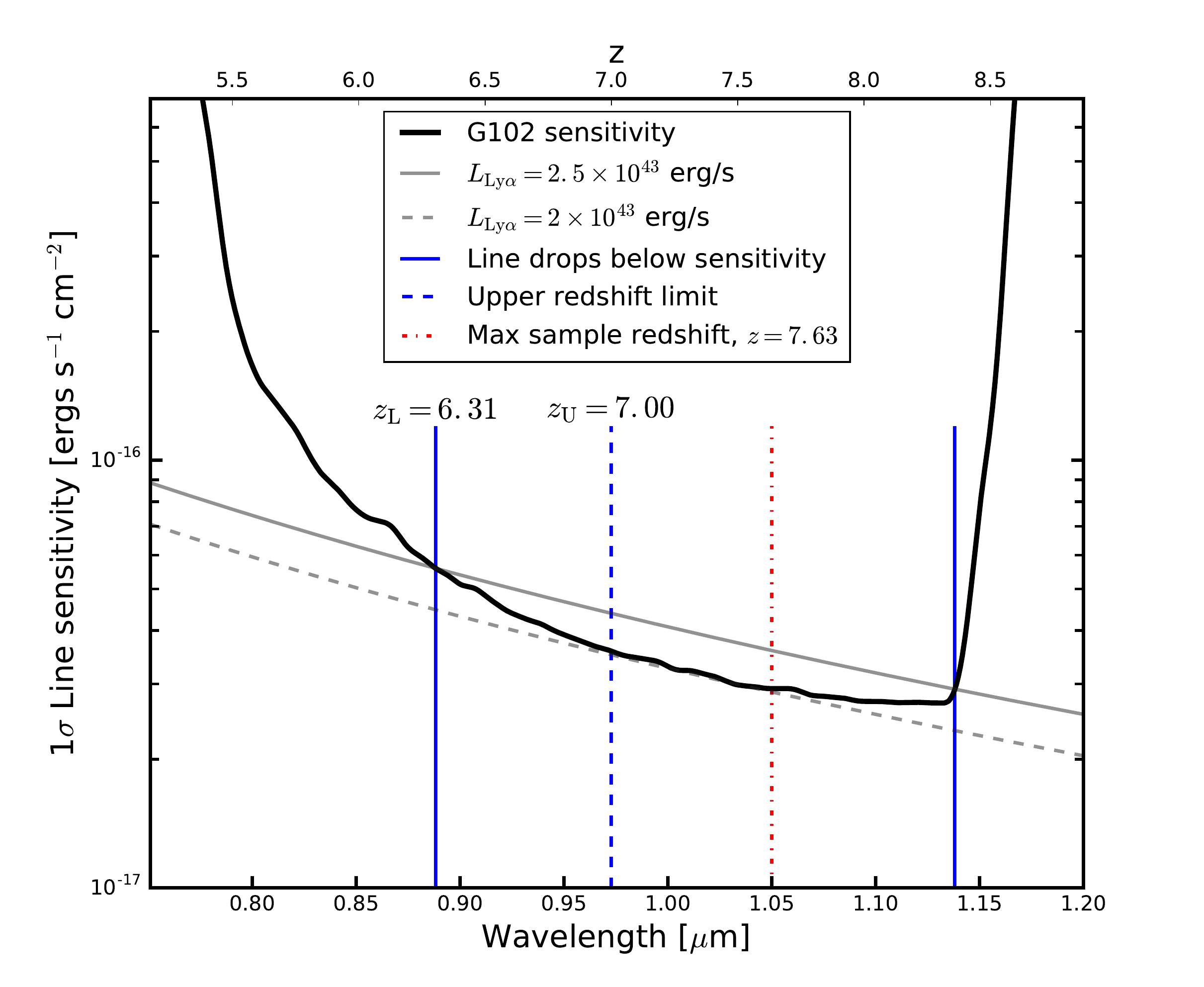}
\caption{Schematic representation of the modified $V_{\mathrm{MAX}}$ method
used to calculate the volume within which an emission line of a given 
luminosity is detectable by the WISP survey. The black curve is the 
G102 sensitivity limit calculated for one of the fields. The solid grey curve
shows the observed flux of a \lya\ emission line with 
$L_{\mathrm{Ly}\alpha}=2.5\times10^{43}$ erg/s at a range of redshifts.
The redshifts at which the emission line drops below the sensitivity curve
are indicated by blue solid lines. In calculating the volume at 
$z\sim6.5$, we set $z_U=7$ (blue dashed line), while the
maximum redshift limit of our sample ($z=7.63$)
is indicated by a red dashed-dotted line. The dashed grey curve shows
the observed \lya\ flux for $L_{\mathrm{Ly}\alpha}=2\times10^{43}$, which
would not be detected in this field.
\label{fig:vol}
}
\end{center}
\end{figure}

\begin{figure}
\begin{center}
\includegraphics[width=0.5\textwidth]{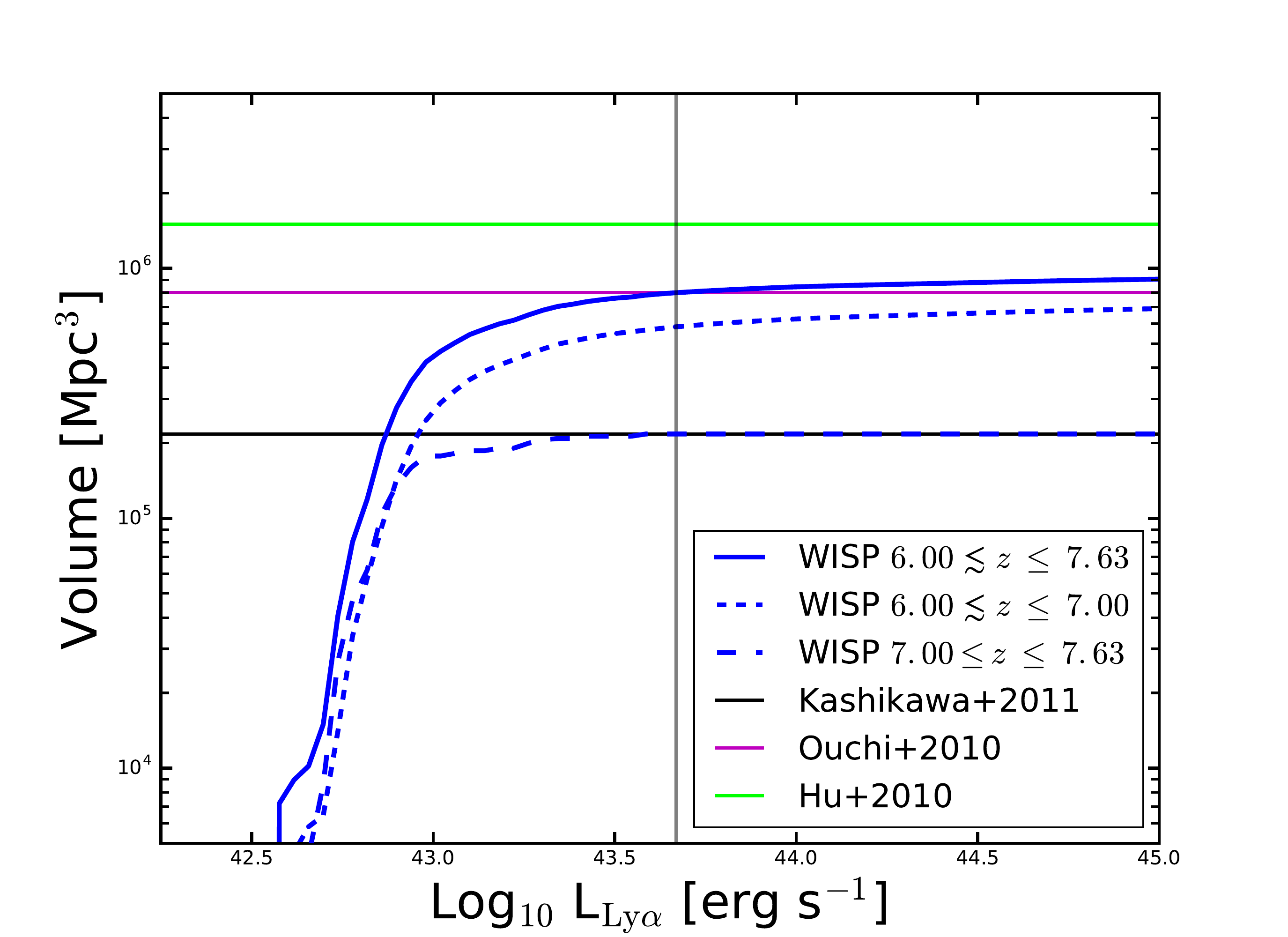}
\caption{The total effective volume reached in all $48$ WISP fields as a
function of luminosity. The solid curve shows the volume probed over the 
full redshift range, $6.00\lesssim z \leq 7.63$. The lower redshift limit 
comes from the drop in sensitivity of the $G_{102}$ grism at the blue end, 
while the upper redshift limit corresponds to our wavelength criterion
$\lambda_{\mathrm{obs,max}} = 1.05 \mu$m.
The dashed curve shows the volume probed from
$6\lesssim z \leq 7$, which is used to calculated the number density of WISP
LAEs in Figure \ref{fig:clf}. The dotted curve shows that from 
$7.00\leq z \leq 7.63$, which is used to place a limit on the number of WISP LAEs
at $z> 7$ (see Figure \ref{fig:z7}). 
\label{fig:surveyvol}
}
\end{center}
\end{figure}

Figure \ref{fig:vol} shows a schematic representation of this process. 
As an example, two curves are plotted in grey showing the flux of a \lya\
emission line as it would be observed if the galaxy were placed 
at a range of redshifts. 
The curves show the fluxes corresponding to 
$L_{\mathrm{Ly}\alpha} = 2\times10^{43}$ erg s$^{-1}$ (dashed) and 
$L_{\mathrm{Ly}\alpha} = 2.5\times10^{43}$ erg s$^{-1}$ (solid).
In the first case, the emission line is too faint to be detected in this
field at any redshift. 
In the second case, the redshifts at which the flux would drop below the 
grism sensitivity are marked by solid blue lines.
The redshift limits in this field would be $z_L=6.31$ and $z_U=7.00$,
where we fix $z_U$ to calculate the effective volume at $z\sim6.5$.
For the volume calculation at $z\geq7$, $z_L=7$ and $z_U=7.63$.
Figure \ref{fig:surveyvol} shows the total effective volume calculated in 
all $48$ WISP fields as a function of line luminosity. 
The total volume probed in WISP fields at the luminosity of the LAEs
is $5.8\times10^5$ Mpc$^3$ between $6\lesssim z \leq 7$. 
For comparison, the volumes covered by the $z\simeq6.5$ 
ground-based narrowband surveys we consider in this paper 
are $1.5\times10^6$ Mpc$^3$ \citep{hu2010},
$8\times10^5$ Mpc$^3$ \citep{ouchi2010}, $2.17\times10^5$ Mpc$^3$ 
\citep{kashikawa2011}, and $42.6 \times 10^5$ Mpc$^3$ \citep{matthee2015}.
Extending the redshift range out to $z=7.63$ would add another 
$\sim2.2\times10^5$ Mpc$^3$ to the WISP volume.

\begin{figure}
\begin{center}
\includegraphics[width=0.5\textwidth]{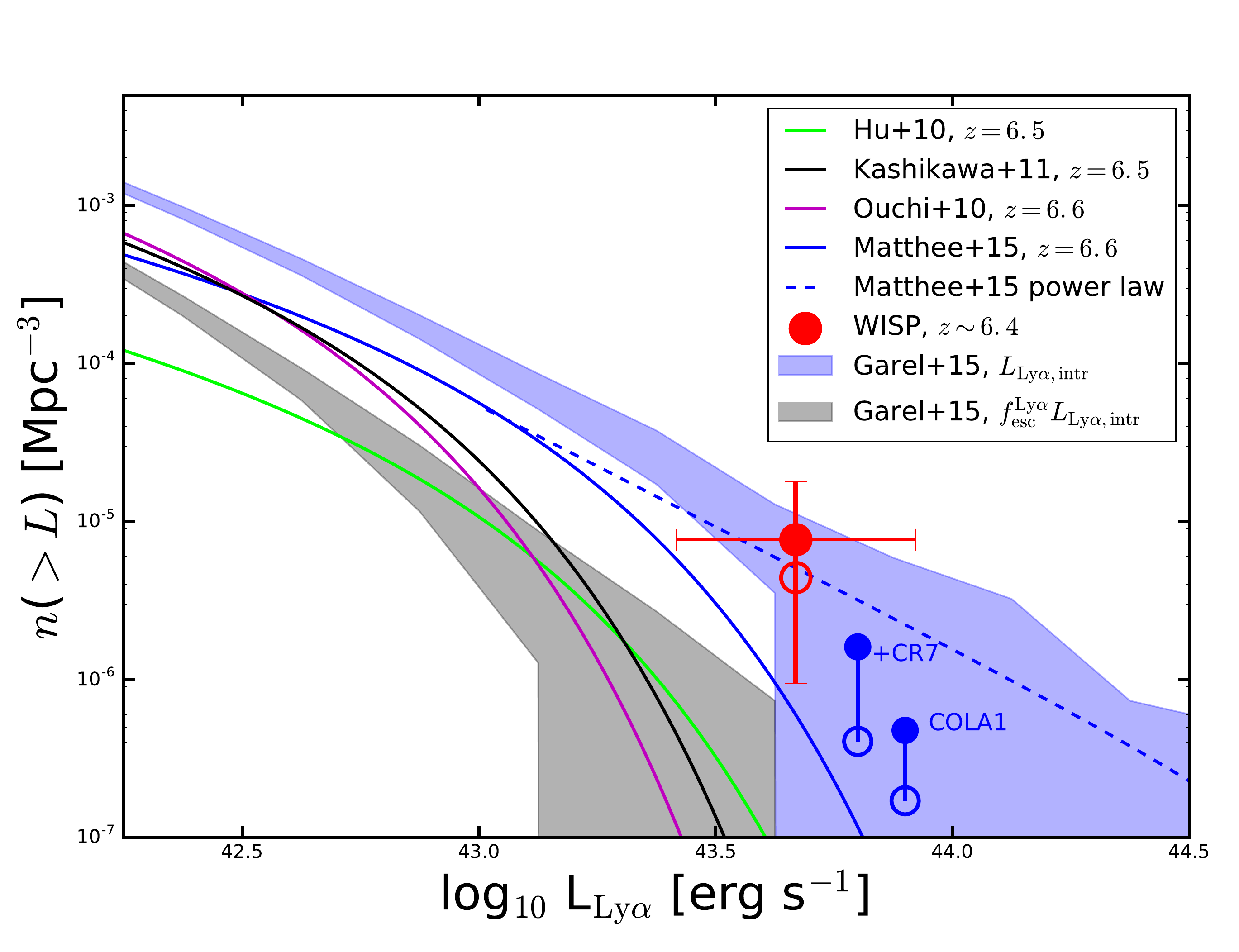}
\caption{The number density of WISP LAEs at $z\sim6.5$. The WISP number 
densities are plotted in red, both with (solid circle) and without
(open circle) completeness corrections. The observed (open blue circles)
and corrected (solid blue circles) number densities calculated 
using the brightest LAEs discovered in COSMOS, CR7 and COLA1, are shown for
comparison. The cumulative LFs of \cite{hu2010}, \cite{kashikawa2011}, 
\cite{ouchi2010} and \cite{matthee2015} are plotted as solid lines. 
The dashed blue line shows the integration of the power law fit to the 
bright end of the LF from \cite{matthee2015}.
The shaded regions show the $1\sigma$ dispersions of the cumulative LF 
calculated from 100 realizations of the mock light cones from \cite{garel2015}.
The light blue and grey regions show, respectively, 
the results using the intrinsic \lya\ 
emission and that which is able to escape.
\label{fig:clf}
}
\end{center}
\end{figure}

We compare the number density of $z\sim6.5$ LAEs in the WISP survey
to measurements of the \lya\ cumulative LF in Figure \ref{fig:clf}.
The completeness-corrected number density calculated according to 
Equation (\ref{eqn:vmax}) is plotted as the solid red circle.
The observed number density -- calculated without the completeness
correction -- is plotted as the smaller, open red circle.
The uncertainty on $n$ is plotted as the Poissonian error for small 
number statistics, and is taken from \cite{gehrels1986}. 
The uncertainty on $log (L_{\mathrm{Ly}\alpha})$ is dominated by the
large flux uncertainty of $\textit{WISP302}$.

The WISP LAEs are among the brightest discovered at these redshifts.
Recently a handful of comparably bright LAEs have been detected and 
spectroscopically
confirmed in the COSMOS field. The two brightest, CR7 
\citep{sobral2015,matthee2015} and COLA1 \citep{hu2016} 
are plotted here in blue.
The number density for CR7 is calculated using the full volume that
\cite{matthee2015} probe in the UDS, COSMOS and SA22 fields,
$42.6\times10^5$ Mpc$^3$. \cite{hu2016} cover an additional 
$\sim2$ deg$^2$ in COSMOS, which we roughly convert to a volume 
using the width of the narrowband filter $NB921$ the authors used to discover
the bright LAEs. COLA1 was discovered in this additional area, which
probed a combined volume of $\sim58.6\times10^5$ Mpc$^3$. 
In each luminosity bin, the open blue circles show the observed number
densities. The closed blue circles are the number densities corrected for 
the shape of the $NB921$ filter profile using the luminosity-dependent 
correction factors presented in \cite{matthee2015}.

We plot the $z\sim6.5$ cumulative LFs of \cite{hu2010}, \cite{ouchi2010},
\cite{kashikawa2011}, and \cite{matthee2015} in Figure \ref{fig:clf}
and discuss this comparison in Section \ref{sec:n65}.

\subsection{Extended Lya emission}
LAEs are expected to have extended \lya~emission. 
\lya~photons created in star-forming regions are resonantly scattered 
outwards, creating large diffuse halos of extended \lya~emission
\citep{zheng2010}. 
At $z\geq6$, where the surrounding IGM is partially
neutral, the \lya~halo could extend as far as 1 Mpc from the galaxy
\citep{zheng2011}. 
Evidence for extended \lya~halos has been detected around galaxies
at $z\sim0$ \citep{hayes2014}, 
$z\sim2-3$ \citep[e.g.,][]{steidel2011,matsuda2012},
and $z=3-6$ \citep{wisotzki2016}. 
At $z\geq6$, these analyses are 
incredibly difficult to perform on individual galaxies and 
narrowband images are almost always stacked \citep{momose2014}.

In the two-dimensional WISP spectra extracted from the full grism images,
spatial information is preserved along the cross-dispersion axis.
We can therefore measure the spatial extent of \lya\ emission
around the LAEs in our sample. 
We create stamps of the 2-D spectra around the emission line of each 
LAE. 
There is no continua detected in the spectra, but to be sure we
fit the background row by row on either side of the emission line 
and subtract it out of the spectral stamp.

\begin{figure}
\begin{center}
\includegraphics[width=0.5\textwidth]{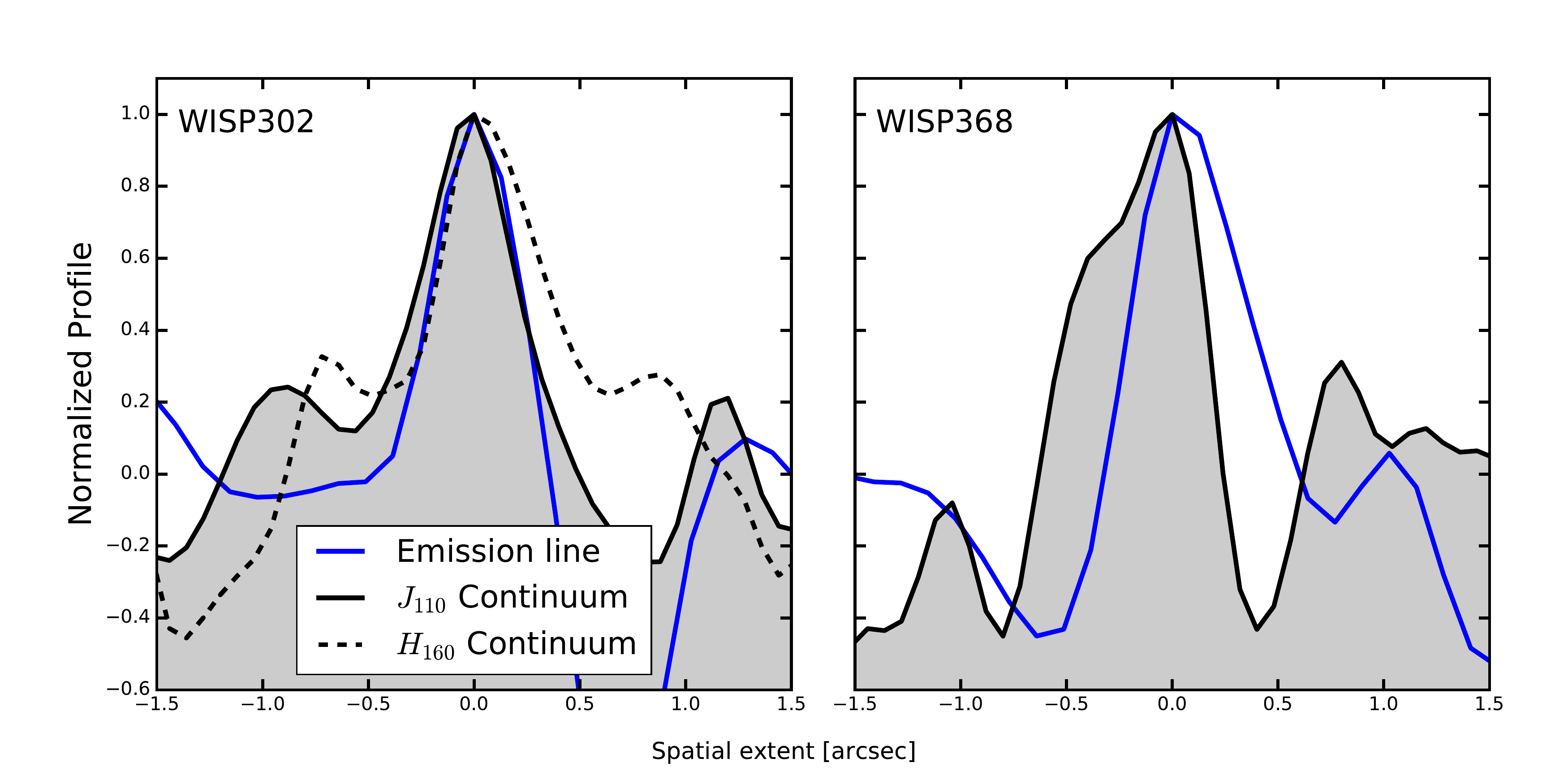}
\caption{The spatial profile of emission from the LAEs.
The \lya\ profile is measured in the 
two-dimensional spectral stamps and is plotted in blue. 
The solid and dashed black curves show the extent of the continuum 
emission measured in the $\jband$ and $\hband$ image stamps, respectively.
Both the \lya\ and continuum emission are compact.
The \lya\ spatial profiles have FWHM$=0\farcs38$ (left, \textit{WISP302}) and
$0\farcs46$ (right, \textit{WISP368}). 
There is a slight offset ($\sim0\farcs3$ measured at half-maximum)
between the \lya\ and continuum profiles of \textit{WISP368}.
\label{fig:ext}
}
\end{center}
\end{figure}
In each stamp, four columns of pixels ($\sim 100$\AA) 
centered on the emission line are collapsed along the wavelength 
direction. This results in a one-dimensional 
profile of the \lya\ emission along the spatial axis as shown in 
blue in Figure \ref{fig:ext}.
In the same way we collapse the continuum image of the galaxy in 
$\jband$ and $\hband$ along the same axis and plot these profiles
in solid and dashed lines, respectively. As \textit{WISP368} is not 
detected in $\hband$, we plot only the $\jband$ profile for this LAE. 
All profiles are normalized to the peak values for easy comparison of the 
profile shapes. 

We measure \lya\ and continuum emission that are both equally compact, with 
full widths at half maximum (FWHM) of $\sim0\farcs4$.
This corresponds to $\sim2.2$ kpc at these redshifts.
Unfortunately, our measurements are limited by the depth of the 
WISP spectra. 
Extended \lya\ emission may be present below our surface brightness limit. 
We discuss this possibility in Section \ref{sec:extem}.

\section{Discussion}\label{sec:discussion}

\begin{figure}
\begin{center}
\includegraphics[width=0.5\textwidth]{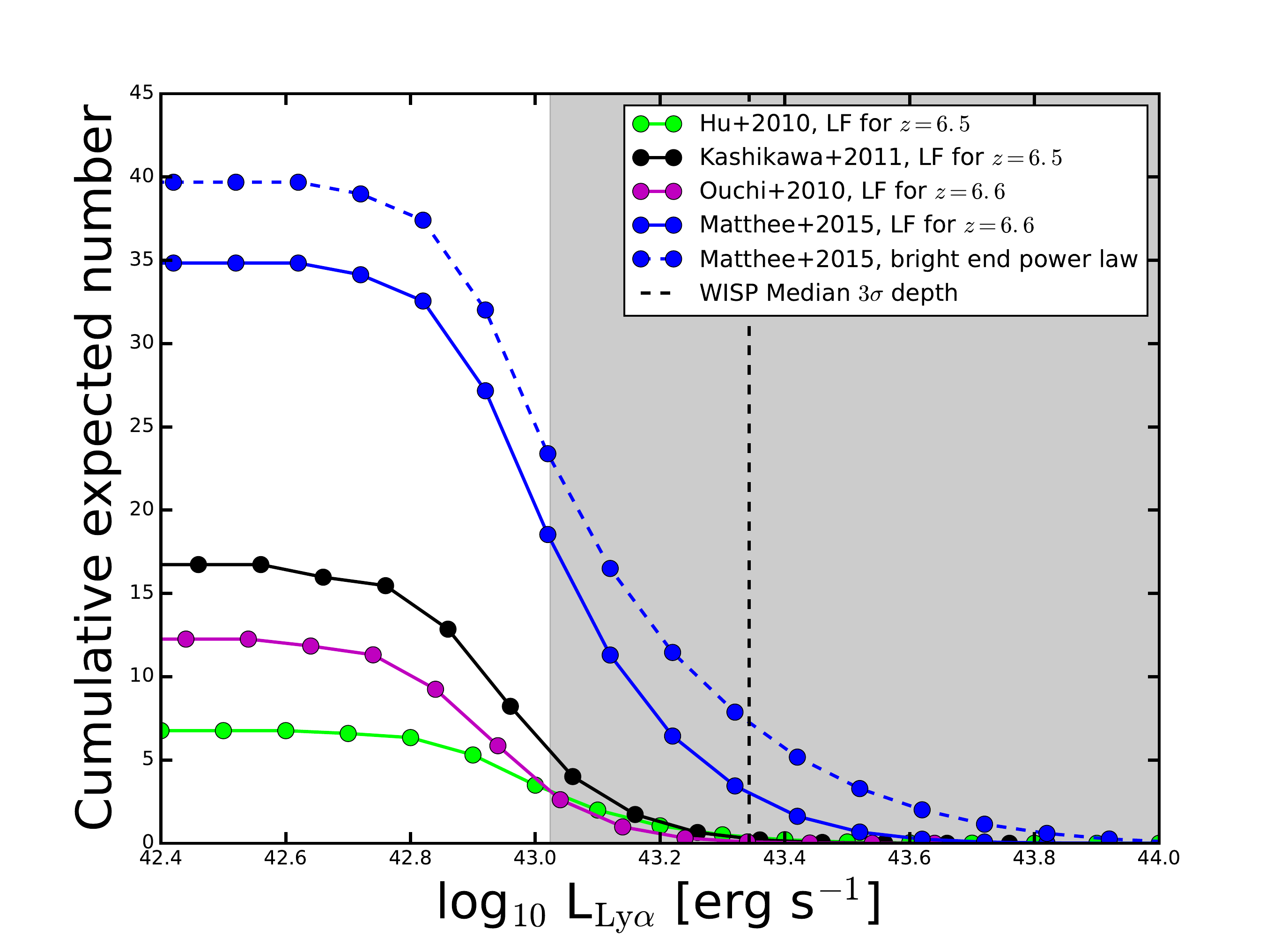}
\caption{Expected number counts of LAEs at $z>7$ assuming no evolution in 
the \lya\ LF from $z\sim6.5$. 
The shaded region indicates the range of $3\sigma$ depths in the 
WISP fields. The median $3\sigma$ depth 
is indicated by the vertical dashed line. 
Based on the four LFs from  
Figure \ref{fig:clf}, we would expect to detect between $\sim0-3$ LAEs in the
volume probed by the WISP survey at $7\leq z \leq7.63$. 
We can rule out evolution in the \lya\ LF of \cite{matthee2015}, but are
unable to make a similar conclusion about the other three LFs.
\label{fig:counts}
}
\end{center}
\end{figure}

\subsection{The number density of $z\sim6.5$ LAEs}\label{sec:n65}
The WISP survey is probing the most luminous LAEs 
at $z\geq6$, a population that can provide important information 
about the state of the IGM at these redshifts. 
It is expected that the observed number density of LAEs decreases as the 
volume-averaged neutral hydrogen fraction of the IGM, $x_{\mathrm{HI}}$, 
increases.
Upon encountering neutral hydrogen, photons at the \lya\ resonance are
scattered out of the line of sight. 
Drops in the observed number densities of LAEs from $z\sim5.5-7$
\citep[e.g.,][]{hu2010,ouchi2010,kashikawa2011} 
are often interpreted as evidence of an increasingly neutral
IGM and indirect measurements of the end of reionization. 

\lya\ photons can escape from the surroundings of a galaxy during the epoch of 
reionization if the galaxy lies in an ionized bubble large enough to
allow the photons to redshift out of resonance before encountering 
the neutral IGM at the edge of the \hii\ region.
We can expect, then, to preferentially observe the most luminous LAEs -- those 
capable of ionizing the largest 
bubbles -- at earlier
times during reionization \citep[e.g.,][]{matthee2015,hu2016}. 
As $x_{\mathrm{HI}}$ increases with redshift, \lya\ 
emission from fainter galaxies -- those that cannot create bubbles out
to sufficiently large radii -- will be increasingly suppressed.
The more luminous galaxies are already surrounded by ionized media and so are
less affected.

Recent studies have indicated that the observed number density of 
bright LAEs is relatively unchanged from $z\sim6.5$ to $z=5.7$
compared with that of the fainter LAEs \citep{matthee2015,santos2016}.
This effect can be seen as a flattening of the \lya\ LF at the bright
end. 
\cite{matthee2015} find that the bright end of the $z=6.6$ 
\lya\ LF is best fit by a power law, indicating that there are more bright
LAEs than expected for a Schechter-like LF. 
In a continuation of the same survey, \cite{santos2016} find that while
there is a drop in the number densities of faint LAEs from $z=5.7$ to 
$z=6.6$, there is almost no evolution at the bright end. 
This result could indicate that the brightest galaxies already reside in 
ionized bubbles by $z\sim6.5$.
Remarkably, our results are consistent with the 
measurements of \cite{matthee2015} and \cite{santos2016}.

Other studies show the opposite effect. \cite{kashikawa2011} find a 
deficit of bright LAEs at $z=6.5$ as compared with $z=5.7$. 
Additionally, at $z=6.5$, the number densities of bright LAEs measured by 
\cite{hu2010}, \cite{ouchi2010} and \cite{kashikawa2011} 
are all far below those of the WISP fields (see Figure \ref{fig:clf}).
For example, \cite{kashikawa2011} find only one LAE with 
$L_{\mathrm{Ly}\alpha} > 10^{43}$ \es.
This result, however, may be heavily influenced by cosmic variance.
While \cite{kashikawa2011} survey a much larger area ($\sim900$ sq. arcmin)
than that presented here ($\sim160$ sq. arcmin), they cover only a single
pointing in a narrow redshift range.
Meanwhile, the $48$ pure parallel WISP fields presented in this paper are
completely uncorrelated. 
We estimate that the cosmic variance in the WISP sample is $<1$\%
\citep{trenti2008}, compared with the $\sim30$\% of \cite{kashikawa2011}. 
The $24-34$\% decrease \cite{kashikawa2011} find in the number density
of LAEs from $z=5.7$ to $z=6.5$ may be due in part to cosmic variance.
However, at log$(L_{\elya}) \sim 43.5$, \cite{matthee2015} observe 
a number density that is $\sim100$ times higher than that of 
\cite{kashikawa2011}, a difference that cannot be explained by 
cosmic variance alone.
Enhanced clustering of LAEs introduced by 
reionization \citep[e.g.,][]{mcquinn2007} could explain 
some of this observed difference.

We note that the surveys presented in Figure \ref{fig:clf} 
have different EW limits, a situation that can affect the measured 
LFs \citep[see, for example, the SED models presented by ][]{konno2016}.
However, \cite{ouchi2008} show through Monte Carlo simulations that 
when they consider all LAEs down to a rest frame EW of $EW_0=0$\AA,
the normalization of the Schechter LF, $\phi^*$, increases by 
$\lesssim10$\% compared with that for EW limits of $EW_0 \sim30-60$\AA.
This exercise indicates that the $\sim5-10$\AA\ difference between the 
EW limits of the WISP Survey and those of, e.g., \cite{ouchi2010} and 
\cite{kashikawa2011} will have at most a minor effect on the 
number density of LAEs we observe.
Additionally, given their fluxes and EWs, 
the WISP LAEs and those discovered by \cite{matthee2015} at the 
bright end would have been easily detected in 
the narrowband surveys of \cite{ouchi2010} and \cite{kashikawa2011}.

In Figures \ref{fig:clf} and \ref{fig:z7}, 
we also compare our results to predictions from 
mock lightcones based on the model of \cite{garel2015} and adapted for the 
area and redshift range covered by the WISP fields. \cite{garel2015} combine 
the GALICS semi-analytic model with numerical simulations of \lya\ radiation 
transfer through spherical, expanding shells of neutral gas and dust 
\citep{verhamme2006} to predict the emission of \lya\ photons in high-redshift 
galaxies and their transfer in galactic outflows, ignoring the effect of IGM 
attenuation. GALICS describes the formation and the evolution of galaxies 
within dark matter halos extracted from a large cosmological simulation box 
($L_\mathrm{box} = 100h^{-1}$ comoving Mpc). For each model galaxy, 
\cite{garel2015} use scaling relations to connect the expansion velocity, 
column density and dust opacity of the shell to the galaxy properties output 
by GALICS \citep[see][for more details]{garel2012}. The \lya\ line profiles 
and escape fractions are then estimated using the library of \lya\ transfer 
models in shells presented in \cite{schaerer2011}.

The shaded regions in Figure \ref{fig:clf} are the $1\sigma$ dispersion 
of the LF measured over 100 realizations of the mock lightcones. The grey 
region, labeled $f_{\mathrm{esc}}^{\elya} L_{\elya \mathrm{,intr}}$, 
shows the LF dispersion for mock lightcones in 
which $f_{\mathrm{esc}}^{\mathrm{Ly}\alpha}$ is calculated as 
described above. We see that this model is in better agreement with the data 
of \cite{hu2010}, \cite{ouchi2010} and \cite{kashikawa2011}, but it cannot 
reproduce the high number density of bright LAEs 
(log$(L_{\mathrm{Ly}\alpha}) \gtrsim 43.5$) found in the WISP survey or by 
\cite{matthee2015}.

We see from Figure \ref{fig:clf} that the dispersion of the LFs estimated 
from the mock lightcones is significant at the bright end, which provides hints 
to the cosmic variance expected in our survey. 
However, the $1\sigma$ dispersion from 
the mock lightcones 
cannot be reconciled with our data point even within the error bars, 
so it seems unlikely that field-to-field variation is responsible for the 
difference between our LF measurement and the LF estimated from the mock lightcones or 
from the other surveys. Nevertheless, we note that cosmic variance is 
underestimated in the mock lightcones of \cite{garel2015} because of the 
finite volume of the simulation box they use, which misses the fluctuations 
on the very large scales.

Another interpretation of the difference between our measurements and the 
predictions from \cite{garel2015} could arise from the fact that they do not 
account for the growth of the HII bubbles during the epoch of reionzation 
when dealing with the \lya\ transfer. Instead, for all model galaxies, \lya\ 
photons need to travel through outflows of neutral gas and dust (as described 
above) which unavoidably reduce their observed luminosities.
Interestingly, we find that our observed number density lies close to the 
mean LF of the \cite{garel2015} models in which $f_\mathrm{esc,\elya} = 1$ 
(the blue region, labeled $L_{\elya \mathrm{,intr}}$, in Figure \ref{fig:clf}). 
This might suggest that most \lya\ photons can easily escape the bright 
WISP LAEs, 
and that there is very little neutral hydrogen in their surrounding medium 
(see also Section \ref{sec:bubbles}).

The observed number density of the WISP LAEs is
consistent with the density of the 
brightest LAEs detected to date at similar redshifts, CR7 and COLA1. 
The WISP number density is also
consistent with that of model galaxies in a completely ionized IGM 
with $f_{\mathrm{esc}}^{\mathrm{Ly}\alpha}=1$.
We expect the WISP LAEs, like CR7 and COLA1, to reside in 
highly ionized bubbles. Such bubbles would enhance the field-to-field
variations in the observed number counts of LAEs and may also partially
explain the differences at the bright end between the LFs in Figure 
\ref{fig:clf}. 
We discuss these bubbles in Section \ref{sec:bubbles}.

If the ionized bubbles are large enough that emission 
blueward of the \lya\ line center is not suppressed by the damping
wings of the neutral IGM,
we may expect the WISP LAE \lya\ line profiles to have blue wings.
Line profiles with both blue and red emission peaks 
are predicted for galaxies with low 
neutral hydrogen column densities \citep[e.g.,][]{verhamme2015}. 
Double-peaked emission is common among 
Green Pea galaxies \citep{henry2015}, and has been tentatively detected
in the spectra of COLA1 \citep{hu2016}.
The detection of blue wings is beyond the resolution of WISP spectra
and is one goal of planned follow-up observations.

As \cite{konno2016} point out, at z=2.2 all LAEs with 
log$(\mathrm{L}_{\mathrm{Ly}\alpha}) \gtrsim 43.4$ may be AGNs.
High-ionization emission lines such as C{\small IV}$\lambda1549$, 
He{\small II}$\lambda1640$ and C{\small III}$\lambda1909$
can be useful in identifying AGNs.
However, these strong UV nebular emission lines 
may also indicate galaxies with young, metal-poor stellar 
populations and large ionization parameters 
\citep{panagia2005,stark2014,stark2015,stark2016}.
At the redshifts of the WISP LAEs,
C{\small IV}$\lambda1549$ falls on the overlap region between $G_{102}$ 
and $G_{141}$, where the sensitivity in the WISP spectra decreases 
significantly. We instead measure upper limits for 
He{\small II}$\lambda1640$ and C{\small III]}$\lambda1909$ fluxes (see 
Table \ref{tab:cand}). 
The limits on the line ratios with respect to \lya\ we measure  
are too uncertain to 
allow us to determine the dominant source of photoionization, and 
we cannot distinguish between AGN, metal-poor galaxies and normal 
star-forming galaxies \citep{stark2014,schaerer2003}.
The emission lines we detect are 
unresolved and therefore the FWHM is less than $\sim1500$ km/s. 
Although we can rule out broad-line AGN, this limit does not allow us to 
distinguish between narrow-line AGN and star-forming galaxies.
We also notice that the \lya\ rest frame EW for the WISP LAEs is many times 
larger than what \cite{konno2016} observe for AGN of similar UV magnitudes. 
Finally, neither \cite{sobral2015} nor \cite{hu2016} find
strong evidence of AGN emission in CR7 and COLA1 spectra, although 
the possibility of AGN contribution remains.
We cannot draw specific conclusions about the physical characteristics 
of the LAEs from the present WISP data.

\subsection{Evolution to $z\gtrsim7$}\label{sec:laez7}
Moving to higher redshifts, 
\cite{konno2014} find a deficit at all luminosities at $z=7.3$ compared 
to that at $z=6.6$. 
This evolution of the \lya\ LF may be the result of an increasing fraction
of neutral hydrogen in the IGM during reionization. 
The authors also suggest that it could be due to the 
selective absorption of \lya\ photons by neutral 
clumps in otherwise ionized bubbles.  
On the other hand, there have been several bright LAEs spectroscopically 
confirmed at $z\geq7$ 
\citep[e.g.,][]{iye2006,vanzella2011,shibuya2012}.
\cite{robertsborsani2016} recently discovered four galaxies at 
$z\sim7-9$, all of which are emitting \lya\ \citep{stark2016}. 
With fluxes $f_{\elya} \sim 0.7 - 2.5 \times 10^{-17}$ \ecs, these 
$z>7$ galaxies are all fainter than or on par with our detection limits. 

We find no LAEs in the redshift range $7.0 < z < 7.63$ in the volume covered
by the WISP fields, $2.17\times10^{5}$ Mpc$^3$.
Assuming the \lya\ LF does not evolve from $z=6.5$ to $z=7$, 
how many would we expect to observe? 
We calculate the expected number of LAEs at $7.00 \leq z \leq 7.63$
given the four $z\sim6.5$ LFs in Figure \ref{fig:clf} and the volumes probed
by the $48$ WISP fields in this redshift range. 
In Equation (\ref{eqn:vmax}), we limit the redshift range so that
$z_{L,i} \geq 7$ and $z_{U,i}\leq7.63$, covering a volume of 
$2.17 \times 10^5$ Mpc. 
The results are plotted in Figure \ref{fig:counts}. 
The shaded region shows the luminosity range to which the WISP fields are 
sensitive. The edge of the region is plotted at the $3\sigma$ flux limit of
the deepest field. 

At the median $3\sigma$ depth of the WISP fields (dashed black line),
we would expect to detect 3.4 LAEs at $7 \leq z \leq 7.63$ based on the 
\cite{matthee2015} Schechter function LF (blue solid line). 
Given our non-detection at $z\geq7$, we find that the probability that this 
$z=6.6$ LF also applies at $z\geq7$ is 3.3\%. 
If we consider the authors' power-law fit to the bright end (dashed blue line), 
the expected number of $z\geq7$ LAEs increases to 7.9, and the probability 
decreases to 0.037\%.
If either LF from \cite{matthee2015} is representative at $z\sim6.5$,
we detect evolution in the \lya\ LF from $z\sim6.5$ to $z>7$.
However, given the LFs of \cite{ouchi2010}, \cite{hu2010}, and 
\cite{kashikawa2011}, we would expect to detect only $0.1-0.5$ $z\geq 7$ LAEs. 
Our observations are inconclusive if these three LFs
are more indicative of galaxy number densities at $z\sim6.5$.

\begin{figure}
\begin{center}
\includegraphics[width=0.5\textwidth]{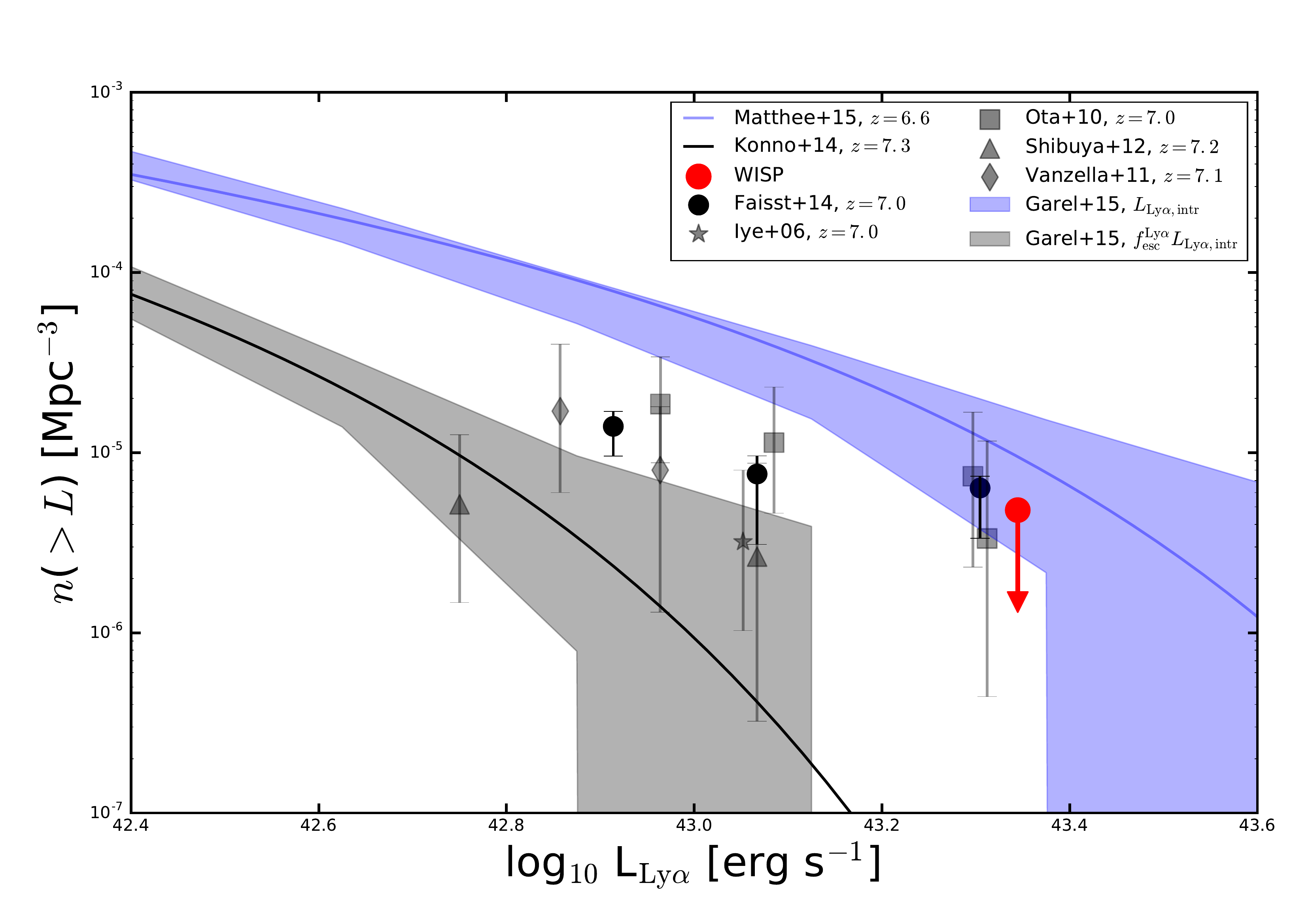}
\caption{Cumulative number densities of LAEs at $z\sim7$. The red arrow 
shows the WISP upper limit calculated from the volume probed over the
redshift range $7 \leq z \leq 7.63$. The grey symbols show the number 
densities of LAEs from other surveys at $z\gtrsim7$. The cumulative
LFs of \cite[$z=7.3$]{konno2014} and \cite[$z=6.6$ for reference]{matthee2015} 
are plotted in black and light blue, respectively. 
The shaded regions are the same as those plotted in Figure~\ref{fig:clf},
but are calculated at these higher redshifts.
\label{fig:z7}
}
\end{center}
\end{figure}
We next use the volume probed at $7 \leq z \leq 7.63$ to place a limit on the
observable, yet undetected 
number density of LAEs over this redshift range. This limit
is shown as a red arrow in Figure \ref{fig:z7} and is calculated at
the median 3$\sigma$ depth of the WISP fields presented in this paper,
$f=3.6\times10^{-17}$ erg s$^{-1}$ cm$^{-2}$.
At $z=7.25$, this corresponds to L$_{\mathrm{Ly}\alpha} = 2.2\times10^{43}$.

The highly uncertain cumulative LF at $z=7.3$ from \cite{konno2014} is
plotted in black. 
The grey symbols show the number densities 
of LAEs at $z\gtrsim7$, including \cite{iye2006} at $z=6.96$;
\cite{ota2010} at $z=7$;
\cite{vanzella2011} at $z=7.008$ and $7.109$; and \cite{shibuya2012}
at $z=7.215$. 
These number densities are as presented by \cite{faisst2014},
where the authors have accounted for the different cosmologies and 
corrections used in the individual surveys. The black circles
show the weighted median that \cite{faisst2014} calculate for these
$z\geq7$ surveys.

The WISP limit is consistent with the other $z\sim7$ measurements. 
Comparing this upper limit with the number density of the WISP $z\sim6.4$ LAEs 
and the $z=6.6$ \cite{matthee2015} LF, we again find 
evidence for a decrease in the number density of LAEs 
from $z\sim6.5$ to $z\gtrsim7$. 
This drop could be due to a higher neutral hydrogen fraction in the
IGM. 
However, it could also at least in part be due to an evolution in the 
intrinsic properties of galaxies at these redshifts, such as
in the escape fraction of \lya\ photons \citep[e.g.,][]{dijkstra2014a}.

As in Figure \ref{fig:clf}, the shaded regions in Figure \ref{fig:z7} 
show the $1\sigma$ dispersions of the \lya\ LFs measured from 
100 realizations of mock lightcones. While still within the 
region corresponding to galaxies with $f_{\mathrm{esc}}^{\mathrm{Ly}\alpha}=1$,
the WISP point is of course an upper limit. 

\subsection{Extent of \lya\ halos}\label{sec:extem}
The detection, or lack thereof, of extended \lya\ emission around a galaxy 
can provide important information about the presence and properties of 
its surrounding neutral hydrogen.
\lya\ photons may escape from a galaxy if they have been redshifted out
of resonance by scattering through a shell of expanding
material \citep[e.g.,][]{verhamme2006}. 
The scattering process acts to diffuse the \lya\ photons
outwards from their point of origin. 
The extent of the resulting \lya\ halo is an indirect probe of the 
fraction of neutral hydrogen present around the galaxy. 
Extended emission is expected to be common around 
galaxies surrounded by partially or mostly neutral hydrogen, while
more compact emission is expected when the surrounding hydrogen 
is ionized (and therefore scattering is minimized).
Extended \lya\ emission has been detected at redshifts ranging from
$z\sim 0.03 - 6$. 
The 14 well-studied galaxies in the \lya\ Reference Sample 
\citep[LARS,][]{hayes2013,hayes2014,ostlin2014} have \lya\ emission that is 
$\sim3$ times more extended than either their \ha\ or FUV emission.
At $z=2.2$, \cite{momose2016} find scale lengths of \lya\ halos
ranging from $7 - 16$ kpc,
while \cite{steidel2011} find evidence for \lya\ halos extending out as 
far as 80 kpc at $z=2.6$.
At $z\sim3.1$, \cite{matsuda2012} detect \lya\ halos out to $\geq60$ kpc.
\cite{wisotzki2016} study a sample of galaxies at $3\leq z \leq6$ and find the 
\lya\ emission is more extended than the UV by a factor of $5-10$.
On the other hand, compact \lya\ emission has been observed at both 
intermediate and high
redshifts: $z\simeq2.1$, $3.1$ \citep{feldmeier2013}; 
$z\simeq5.7$ and $6.5$ \citep{jiang2013}.
The observed compact emission could 
indicate that the galaxies at these redshifts are surrounded
by highly ionized hydrogen, even at the end of reionization at 
$z\sim6.5$.

It is likely, however, that measurements of compact \lya\ emission
such as those of \cite{feldmeier2013} and \cite{jiang2013}
reflect the surface brightness limits of the observations
\citep[e.g.,][]{steidel2011,schmidt2016,wisotzki2016} rather then 
the physical environments surrounding the galaxies.
\cite{guaita2015} redshift LARS galaxies to simulate observations 
of these galaxies at $z\sim2$ and $z\sim5.7$.
By $z=5.7$, the extended \lya\ emission is only detected in 
stacked narrowband images, and the stacked \lya\ profile drops below the surface
brightness limit at $\sim5$ kpc.
Repeating this process for galaxies at $z=7.2$ for
the Grism Lens-Amplified Survey from Space (GLASS), \cite{schmidt2016}
find that the surface brightness of the extended \lya\ emission in LARS
galaxies is too faint
to be detected in their stacked spectra.

Both LARS and GLASS reach depths comparable to or slightly deeper than WISP:
$3\times10^{-18}$ \ecs\ in the $z=5.7$ simulated LARS spectra and
$3.5-5\times10^{-18}$ \ecs\ (observed, uncorrected for magnification) 
in the GLASS spectra.
The WISP spectra are therefore only sensitive to the 
central, high surface brightness core of the \lya\ emission. 
Indeed, the $1\sigma$ $G_{102}$ depths reached in the two WISP fields are 
$5.5\times10^{-17}$ erg s$^{-1}$ cm$^{-2}$ arcsec$^{-2}$, 
a factor of 550 times brighter
than that reached by \cite{wisotzki2016}. 
We therefore do not draw conclusions from the compact \lya\ profiles we 
measure, but instead suggest 
the LAEs presented here are good candidates for targeted, deep follow-up 
observations aimed at exploring the extent of the \lya\ halo at $z\sim6.5$.

\subsection{Ionized bubbles}\label{sec:bubbles}
It is instructive to compare the observed \lya\ luminosities with 
the intrinsic one, computed with basic assumptions on the stellar population 
of the WISP LAEs. 
We consider a range of parameters including both \cite{chabrier2003} 
and \cite{salpeter1955} IMFs, metallicities of $Z/Z_{\odot} = 0.005, 0.02$ 
and $0.2$, and ages from 10 Myr to 1 Gyr.
Assuming Case B recombination,  $n_e = 100$ cm$^{-3}$ and $T_e = 10^4$ K, 
the intrinsic \lya\ luminosity of a galaxy can be estimated from the 
production rate of ionizing photons as \citep[e.g.,][]{schaerer2003}:
\begin{equation}
\lint = C \ (1 - \fesc) \ N\gamma,
\label{eqn:nion}
\end{equation}
where $C = 1.04\times10^{-11}$ erg and $\fesc$ is the escape fraction 
of ionizing photons.
$N\gamma$ can be computed by scaling the stellar population model to the  
observed $\hband$ magnitude (rest--frame 2000\AA, not contaminated by the \lya\ 
emission). For reference, we adopt $\fesc=0$ and perform this calculation for 
\textit{WISP302} only, since \textit{WISP368} is undetected in the $\hband$ 
continuum.

According to this model, and assuming the lowest metallicity 
covered by the BC03 models ($Z=0.0005Z_{\odot}$) and an age of 10 Myr,
\textit{WISP302} can produce $5.1\times10^{54}$ ionizing photons per second.
The ratio of observed to intrinsic \lya\ emission is therefore $\sim 0.9$, 
implying that we are detecting almost all of the \lya\ photons produced 
by this galaxy.
More moderate parameter values result in an observed 
$L_{\mathrm{Ly}\alpha,\mathrm{obs}}$ 
that is greater than the intrinsic $L_{\mathrm{Ly}\alpha,\mathrm{intr}}$.
Only models with a very young age and low metallicity produce a ratio 
$L_{\mathrm{Ly}\alpha,\mathrm{obs}}/L_{\mathrm{Ly}\alpha,\mathrm{intr}} < 1$.

This exercise suggests that these photons are not only able to escape 
from the galaxy's ISM, but they 
are also not substantially attenuated by the surrounding IGM.
In fact, local (i.e., where the IGM is not affecting the \lya\ profile) 
analogs to high-$z$ galaxies with large escape frations of
\lya\ radiation are characterized  by an emission line profile showing 
substantial flux blueward and close to the systemic 
velocity \citep{henry2015,verhamme2015}.  The neutral IGM at $z\sim 6.5$ would 
attenuate this blue emission considerably, causing  
$L_{\mathrm{Ly}\alpha,\mathrm{obs}}/L_{\mathrm{Ly}\alpha,\mathrm{intr}} \lesssim 0.5$.
Hence, we conclude that a sizable neutral fraction in the presence of 
this galaxy is unlikely.
This conclusion is also supported by the agreement between the 
LF calculated from the intrinsic \lya\ emission \citep{garel2015} 
and the high number density of WISP LAEs (see Figure~\ref{fig:clf}).

It is possible that the stellar population of \textit{WISP302} 
is more  extreme than our simple assumption. 
Both Pop~III stars and models that include binaries have effectively 
harder spectra than typical Pop~II stars, and can produce a factor of 
up to $\sim 3$  more ionizing photons than the spectral template discussed
above \citep{schaerer2002,eldridge2008}. 
A larger $L_{\mathrm{Ly}\alpha,\mathrm{intr}}$ would imply a much lower 
$L_{\mathrm{Ly}\alpha,\mathrm{obs}}/L_{\mathrm{Ly}\alpha,\mathrm{intr}}$ ratio 
removing the need for a fully ionized IGM around this object. 
We can search for evidence of extreme stellar populations in \textit{WISP302} looking for the \heii$\lambda1640$\AA\ emission line \citep{panagia2005,stanway2016} as well as other UV nebular emission lines 
\citep[e.g.,][]{stark2014}.

The presence of dust could also reduce the 
$L_{\mathrm{Ly}\alpha,\mathrm{obs}}/L_{\mathrm{Ly}\alpha,\mathrm{intr}}$ ratio. 
If dust were present, $\lint$ would be higher than what we 
estimate from our simple dust--free model. 
At the absolute magnitude of \textit{WISP302}, however, 
\cite{bouwens2016} show that the correction for dust extinction is 
negligible ($< 0.2$ magnitudes).  
As we cannot reliably measure the UV slope 
$\beta$, we rely on this estimate of dust extinction.

If extreme stars are not important, then the high $L_{\mathrm{Ly}\alpha,\mathrm{obs}}/L_{\mathrm{Ly}\alpha,\mathrm{intr}}$ ratio indicates that the IGM is mostly ionized around this galaxy. 
The current measurement of the Thomson optical depth from the cosmic microwave background radiation would allow for an end of the reionization process by as early as $z\sim6.5$ \citep{planck2016,robertson2015}.  However, if this were the case, we would expect bright galaxies such as the WISP LAEs to be more common.  Therefore, it is  more likely that objects like \textit{WISP302} reside in rare, localized bubbles of ionized gas. 

We now investigate the sources that are needed to create a 
sufficiently large ionized bubble for the \lya\ emission to escape the effects of the IGM damping wing.  \lya\ photons will be transmitted through the IGM if the optical depth they experience is $\tau_{IGM} < 1$.  \cite{miraldaescude1998b} shows that the minimum radius required is $R_{\mathrm{min}} \sim 1.216$
proper Mpc.  
Following \cite{cen2000}, 
we can define $R_{\mathrm{max}}$, the maximum radius ionized over the 
course of \textit{WISP302}'s lifetime as:
\begin{equation}
R_{\mathrm{max}} = \left ( \frac{3 \ N_{\mathrm{ion}}}{4 \ \pi \ \langle n_{\mathrm{H}} \rangle} \right )^{1/3}.
\label{eqn:rs}
\end{equation}
Here, $\langle n_{\mathrm{H}} \rangle$ is the mean hydrogen density within
$R_{\mathrm{max}}$ and $N_{\mathrm{ion}}$ is the total number of ionizing 
photons emitted. 
$N_{\mathrm{ion}}$ can be expressed as the product of the rate of ionizing photons produced by the galaxy, the 
fraction of these that escape, and the galaxy's lifetime: 
$N_{\mathrm{ion}} = N_{\gamma} \ \fesc \ t$. 
We assume the same stellar population as above, but we consider a longer star-formation episode (i.e., $t=100$ Myr).
This age will provide us with a conservative upper limit on the radius of the ionized region. 

Following \cite{stiavelli}, we calculate the hydrogen number density 
from the mean density of a virialized halo, $\rho_{\mathrm{vir}}$:
\begin{align*}
\rho_{\mathrm{vir}} &= \xi \ \Omega_M \ \rho_0 \ (1 + z)^3 \\
n_{\mathrm{H}} &= \frac{\rho_{\mathrm{vir}} \: \Omega_b}{\mu \: m_{\mathrm{p}} \: \Omega_M}.
\numberthis \label{eqn:nh} \\
\end{align*}
where $\xi \simeq 178$ is the ratio between the density of a virialized
system and the matter density of the universe;
$\mu\simeq1.35$ is the mean gas mass per hydrogen atom and accounts 
for the contribution of helium; $m_{\mathrm{p}}$ is the mass of a proton;
and we adopt $\Omega_bh^2 = 0.0223$ \citep{planckparameters}.
We find $R_{\mathrm{max}} \simeq 0.45 (f_{esc})^{1/3}$ Mpc. 
Even if all the ionizing photons escaped the galaxy's 
ISM ($f_{\mathrm{esc}}=1$),
the size would still be less than half of the minimum required radius, 
$R_{\mathrm{min}}$.

Clearly, \textit{WISP302} is not capable of ionizing a large enough bubble 
on its own. 
We consider two alternative  sources for the needed ionizing photons: 
a population of faint galaxies, as in, e.g., \citet{vanzella2011} 
and \cite{castellano2016}, 
or a bright quasar close to the studied LAE but outside the WISP field of view.

We begin with the first possibility. 
From Equation (\ref{eqn:rs}) we find that at least
$1.45 \times 10^{57}$ photons/s are required to create a sufficiently large ionized bubble, where we have assumed an escape fraction of $f_{\mathrm{esc}}=0.1$. 
We next determine how many galaxies are needed to produce the required
number of photons.
To do so, we find the \lya\ luminosity density by integrating 
\begin{equation}
\int \ L \ \phi(L) \ \mathrm{d}L,
\label{eqn:lumdensity}
\end{equation}
where $\phi(L) \mathrm{d}L$ is the LF, and normalize by the observed 
number density of WISP LAEs.
We then convert the \lya\ luminosity density to a density of ionizing photons 
using Equation (\ref{eqn:nion}). 
Faint galaxies are the dominant source of ionizing photons, and so we assume a 
simple power law LF and begin by considering a slope of $\alpha=-1.5$, 
matching the slopes plotted in Figure \ref{fig:clf}.
In order to reach the requisite $10^{57}$ photons/s, we must integrate
Equation (\ref{eqn:lumdensity}) down to 
$L_{\mathrm{Ly}\alpha} \sim0.001 L^*_{\mathrm{Ly}\alpha}$.
This luminosity limit is on par with the
minimum UV luminosities adopted by, e.g., \cite{robertson2015} and 
\cite{rutkowski2016}, in evaluating the contribution of star-forming 
galaxies to reionization. 
If we allow for a steeper faint-end slope such as $-2.3 < \alpha < -2.0$
\citep[e.g.,][]{dressler2015}, we need only integrate
down to $L_{\mathrm{Ly}\alpha} \sim 0.04 - 0.1 L^*_{\mathrm{Ly}\alpha}$. 
A substantial, but not unreasonable, 
number of faint galaxies is required to produce $10^{57}$ ionizing photons 
in the volume. 
Alternatively, in the presence of an overdensity, an increase 
in the luminosity function normalization could further relax the need for 
faint galaxies.
These calculations strongly depend on the escape fraction of 
Lyman continuum photons
in these high-$z$ galaxies \citep[e.g.,][]{rutkowski2016,vanzella2016,smithbm2016}. 

We now consider the second possibility that  \textit{WISP302}  resides in a bubble ionized by 
a nearby quasar.  For the same UV luminosity, quasars produce approximately an order of magnitude more  
 ionizing photons than star-forming galaxies. Moreover, the escape fraction of ionizing radiation is likely 
close to 100\% \citep[e.g.,][]{loeb2001,cristiani2016}. 
For a single QSO to produce the required number of ionizing photons, 
it would need to be as bright as $\hband \sim 22$, not unreasonable 
given the luminosities of $z>6$ QSOs observed by \cite{fan2006}.
This object would be easily identified in the WISP survey, although none is detected in the region observed around \textit{WISP302}.
At $z=6.4$, however, one Mpc corresponds to $\sim 3^{\prime}$. 
Thus, it is possible that the quasar falls outside of the WFC3 field. 
At this redshift, the average baryonic density of the universe is such that the 
recombination time is of the order of the age of the universe, 
and shorter than typical QSO lifetimes \citep{trainor2013}. Once 
ionized, the bubble could then remain ionized after the QSO has turned off.
Imaging and spectroscopy over a wider area around \textit{WISP302} are needed 
to investigate these possibilities.

\section{Summary}
In this paper we present the results of a search for $z>6$ LAEs in the WISP
survey.  We find two $z\sim6.5$ LAEs in $\sim160$ sq. arcmin, probing a volume of $5.8\times10^5$ Mpc$^3$.
We estimate the contamination fraction in our sample selection to be $\lesssim2$\%, owing mainly to rare \oiii-emitters at $z\sim 0.8$
with \ha/\oiii\ $< 0.25$.

Our number density of bright WISP LAEs at $z\sim6.5$  is higher than the previously reported measurements by \cite{hu2010}, 
\cite{ouchi2010},
and \cite{kashikawa2011}, and is consistent with the findings of
\cite{matthee2015}. 
The discrepancy between the WISP measurement and these studies is likely due to a combination of our larger redshift coverage
(and therefore comparable or even larger volume) and  the almost negligible effect of cosmic variance in our sample.

We do not detect any $z>7$ LAEs, and determine that the $z=6.6$ \lya\ LF as measured by \cite{matthee2015} must evolve
 from $z=6.6$ to $z>7$. 
Based on the expected number counts at the median depth of our fields,
we cannot make a similar claim for the other $z\sim6.5$ LFs.
Our upper limit on the observed number density of LAEs in the WISP volume 
from $7 \leq z \leq 7.63$ is consistent with other $z\sim7$ measurements.

We argue that the WISP LAEs reside in large ionized bubbles in the IGM. 
Using simple assumptions on their stellar populations, we conclude 
that they are not capable of ionizing their surroundings. 
We suggest that either a nearby bright quasar or a substantial 
population of fainter galaxies is required. 

The WISP LAEs are excellent targets for  studies of the IGM towards the end of reionization as well as the sources
contributing to the ionizing photon budget. Follow-up observations can explore whether there is evidence 
for extreme stellar populations in these galaxies by 
targeting the \heii\ emission line as well as other UV nebular 
emission lines \citep[e.g.,][]{stark2014}.
We can look for the presence of a blue wing in the \lya\ line profile,
and obtain deep spectroscopic observations to investigate the absence (presence) of low 
surface brightness components around our LAEs to support (refute) the idea that 
these sources are located in ionized bubbles.
Measurements of the LAEs' systemic velocities are also needed 
to determine whether the \lya\ emission is close to the systemic velocity, 
as observed in some local galaxies with large \lya\ escape fractions, or 
substantially redshifted. We note, however, that both the ISM and IGM can contribute to the redshifting of the line, and these contributions cannot be separated at these high redshifts.    

We can observe a wider field-of-view to search the surrounding volume
for  bright quasars, or probe down to fainter \lya\ luminosities to study the 
population of fainter galaxies around the bright LAEs.
These observations can provide information on the sources 
mainly responsible for ionizing the regions around the LAEs.
Follow-up observations can confirm whether there really are sufficient  
galaxies to create the bubbles, or whether a more powerful source of ionizing photons is required.

Finally, upcoming space-based dark energy missions have the 
potential to discover thousands of these bright LAEs. 
For example, covering 40 sq. degrees at a $3.5\sigma$ sensitivity of 
$6 \times 10^{-17}$ erg s$^{-1}$ cm$^{-2}$, 
the Euclid Deep Survey \citep{laureijs2012} has the potential to 
identify $\sim70$ LAEs per sq. degree at $6.5\lesssim z \lesssim 7.0$.
These targets would be ideal for ground-based and JWST follow-up.

\acknowledgements
We thank the anonymous referee for a careful review and suggestions
that improved the manuscript.
This research was partially supported by a Jet Propulsion Lab/NASA grant
to the University of Minnesota (RSA 1541900 / NNN12AA01C).
Support for HST Programs GO-11696, 12283, 12568, 12902, 13517, 13352, and 14178
was provided by NASA through grants from the Space Telescope Science 
Institute, which is operated by the Association of Universities for 
Research in Astronomy, Inc., under NASA contract NAS5-26555.
TG is grateful to the LABEX Lyon Institute of Origins 
(ANR-10-LABX-0066) of the Universit\'e de Lyon for its financial support 
within the programme `Investissements d'Avenir' (ANR-11-IDEX-0007) of the 
French government operated by the National Research Agency (ANR).
This research has made use of NASA's Astrophysics Data System
Bibliographic Services.

\bibliography{bagley_accepted}

\appendix
\section{Simulated template library}\label{sec:library}
For the purposes of deriving the optimal color selection for $z\geq6$ galaxies,
we generate a catalog of $20000$ synthetic sources based on the 
models of \citet[hereafter BC03]{bc03}.
The catalog will be used to determine the broadband colors of galaxies
with a variety of properties, and so we will be calculating the mean flux
density of each template spectrum in the four WFC3 filters used in this paper.
Strong emission lines that fall within a filter passband can contribute 
significantly to the photometry, as much as $0.75$ magnitudes or more,
and can therefore have a large impact on broadband colors
\citep[e.g.,][]{atek2011}.
In order to account for this effect in our synthetic catalog,
we add emission lines to exactly half of the $20000$ templates.
In what follows, we describe the templates and their properties in more 
detail. Unless otherwise noted, all properties -- redshift, metallicity, 
star formation history, dust content, etc. -- are drawn randomly from uniform 
distributions.

The template redshifts range from $0.1 \leq z \leq 8.5$. 
The luminosities of the synthetic sources in each redshift bin
are distributed according to the LF at that redshift.
To achieve this, 
in each of eight redshift bins,
we integrate the appropriate LF down to a luminosity that corresponds to the 
$1\sigma$ magnitude limit of our survey and assign each source a luminosity
extracted from this cumulative LF. 
Table \ref{tab:lfs} lists the LFs used in each redshift bin.

\begin{table*}
\begin{center}
\begin{threeparttable}
\caption{Luminosity Functions used in synthetic catalog}
\label{tab:lfs}
\begin{tabular}{@{}lccccc}
\toprule
& Redshift & $M^*_{UV}\footnotemark$  & $\phi^* (10^{-3}$ Mpc$^{-3})$ &         $\alpha$ & Reference \\
\midrule
& $0.1\leq z<0.6$ & $-21.03\pm0.25$ & $5.0\pm0.6$  & $-1.26\pm0.15$ &           \cite{scarlata2007}  \\
& $0.6\leq z<1.5$ & $-21.24\pm0.12$ & $4.9\pm0.3$  & $-1.22\pm0.10$ &           \cite{scarlata2007}\footnotemark  \\
& $1.5\leq z<2.7$ & $-20.01\pm0.24$ & $2.54\pm0.15$ & $-1.74\pm0.08$ &          \cite{alavi2014} \\
& $2.7\leq z<3.4$ & $-20.97\pm0.14$ & $1.71\pm0.53$ & $-1.73\pm0.13$ &          \cite{reddy2009} \\
& $3.4\leq z<4.5$ & $-20.88\pm0.08$ & $1.97^{+0.34}_{-0.29}$  & $-1.64\pm0.04$ &           \cite{bouwens2015} \\
& $4.5\leq z<5.5$ & $-21.17\pm0.12$ & $0.74^{+0.18}_{-0.14}$  & $-1.76\pm0.05$ &   \cite{bouwens2015} \\
& $5.5\leq z<6.5$ & $-20.94\pm0.20$ & $0.5^{+0.22}_{-0.16}$  & $-1.87\pm0.10$ &   \cite{bouwens2015} \\
& $6.5\leq z<7.5$ & $-20.77\pm0.28$ & $0.34^{+0.24}_{-0.14}$  & $-2.03\pm0.13$  & \cite{bouwens2015} \\
& $7.5\leq z\leq8.5$ & $-20.21\pm0.33$ & $0.45^{+0.42}_{-0.21}$  & $-1.83\pm0.  25$ & \cite{bouwens2015} \\
\bottomrule
\end{tabular}
\begin{tablenotes}
  \small
  \item $^{\textrm{a}}$ Values of $M^*_{UV}$ are derived at the rest
    frame wavelengths of the $B$-band $4420${\AA}
    for $0.1\leq z<1.5$, $1500${\AA} for $1.5\leq z <2.7$, 
    $1700${\AA} for $2.7\leq z<3.4$, and $1600${\AA} for
    $3.4\leq z \leq 8.5$.
  \item $^{\textrm{b}}$ LF is calculated for $0.6\leq z\leq0.8$ but here has
    been extended down to $z=0.1$ and up to $z=1.5$. 
\end{tablenotes}
\end{threeparttable}
\end{center}
\end{table*}

Each source is then assigned a spectral template extracted from a 
library of BC03 templates. 
We allow for either \cite{chabrier2003} or 
\cite{salpeter1955} initial mass functions, composite stellar 
populations of constant metallicity ($Z/Z_{\odot} = 0.005$, $0.02$,
$0.2$, $0.4$, or $1$),
and exponentially-declining star formation histories with 
characteristic timescales of $\tau = 0.01$, $0.5$, and $5.0$ Gyr.
The timescales considered here approach those of models with, 
at either extreme, a constant SFH (large $\tau=5$) and a single burst of
star formation (small $\tau$). 
We restrict the template metallicity with redshift such that 
the maximum possible metallicity (in units of $Z_{\odot}$) is
$Z\leq0.4$ at $1<z\leq2$, 
$Z\leq0.2$ at $2<z\leq3$, and 
$Z\leq0.02$ for $z>3$. 
The full metallicity range is available for synthetic galaxies at $z\leq1$.
The spectra are extracted from these templates in logarithmic time steps
such that the spectral evolution is better sampled for younger ages. 
Sources are then randomly assigned an age ranging from 10 Myr to the 
age of the universe at the given redshift. 

When adding emission lines to the templates, the line luminosity
for \hb\ is determined from the flux of 
hydrogen-ionizing photons output by the BC03 models. Emission line
coefficients for \ha\ and \lya\ are from Table 1 of \cite{schaerer2003}.
We assume a \lya\ escape fraction of $f_{\mathrm{esc}}^{\mathrm{Ly}\alpha}=1$. 
The ratios of other rest-frame optical emission lines, relative to \hb,
depend on each template's metallicity, the electron temperature and 
density. We assume $n_e=100$ cm$^{-3}$
and $T_e=10000$ K and take the resulting ratios from \cite{anders2003}.

These templates of stellar and nebular emission are next reddened using
two distinct dust geometries \citep[see][]{scarlata2009}:
(1) a uniform slab of dust in front of the source, and 
(2) a clumpy slab in front of an extended source. In the first
case, extinction follows the classical $I_{o}/I_{i} = e^{-\tau_{\lambda}}$,
where $I_{o}$ and $I_{i}$ are the observed and intrinsic intensities, 
respectively, and $\tau_{\lambda} = \kappa_{\lambda} \ E(B-V) / 1.086$.
We use the \cite{calzetti2000} reddening curve for $\kappa_{\lambda}$,
with color excess values in the stellar continuum drawn randomly from a 
uniform distribution between
$0 < E_s(B-V) < 1$. 
For the second model, we assume all clumps have the same optical depths,
$\tau_{c,\lambda}$, and they are distributed according to a Poisson 
distribution with mean equal to the average number of
clumps along the line of sight, $N$. The number of clumps is randomly
drawn from a uniform distribution between $1 \leq N \leq 10$.
We assign each clump an optical depth in the $V$ band such 
that $0.1 \leq \tau_{c,V} \leq 10$.  Each clump acts as a uniform dust
screen and follows the \cite{cardelli1989} extinction law. The ISM
extinction then goes as
$I_{o}/I_{i} = \textrm{exp}[-N(1-e^{-\tau_{c,\lambda}})]$
\citep{natta1984}.
We randomly assign one of the two dust geometries to each template.
Assuming the dust content of the universe decreases exponentially with
increasing redshift, we restrict the maximum available extinction $E_{B-V}$
(either in the uniform ISM or in the clumps) following
\cite{hayes2011}, where $E_{B-V}(z) = C_{EBV} \ \mathrm{exp}(z/z_{EBV})$,
$C_{EBV}=0.386$ and $z_{EBV}=3.42$. 

The reddened spectra are then attenuated by the neutral hydrogen in the
IGM using the recipe from \cite{inoue2014}, which is an updated
version of the \cite{madau1995} attenuation model.
\cite{inoue2014} include absorption from the full Lyman series,
Lyman limit systems and damped \lya~systems, as well as a more
accurate represention of absorption in the Lyman continuum. 

For each simulated galaxy, we compute the $\vband$, $\iband$, $\jband$, and
$\hband$ magnitudes. Finally, we add errors to the synthetic magnitudes
to account for the photometric uncertainties in our observed catalog.
Photometric scatter is one of the dominant causes of contamination 
in Lyman break samples \citep[e.g.,][]{stanway2008}. It operates in both
directions, scattering galaxies both into and out of the selection window,
thus affecting both the contamination and recovery fractions.

We add photometric scatter to our synthetic catalog in the following way.
From the WISP photometric catalog, we determine the average uncertainty 
as a function of magnitude for each filter. The magnitudes of the synthetic
sources are then allowed to vary according to Gaussian distributions with 
standard deviations set by the median uncertainty in the corresponding
magnitude bins. We create $10$ such realizations of the synthetic photometry
resulting in a final catalog of $2\times10^5$ sources. This catalog
is used in Section \ref{sec:criteria} to choose the selection criteria for 
$z>6$ LAEs, and also to characterize possible sources of contamination 
in the sample in Section \ref{sec:contam}. 

\section{Additional sources}\label{app:others}
\begin{figure*}
\begin{center}
\includegraphics[width=\textwidth]{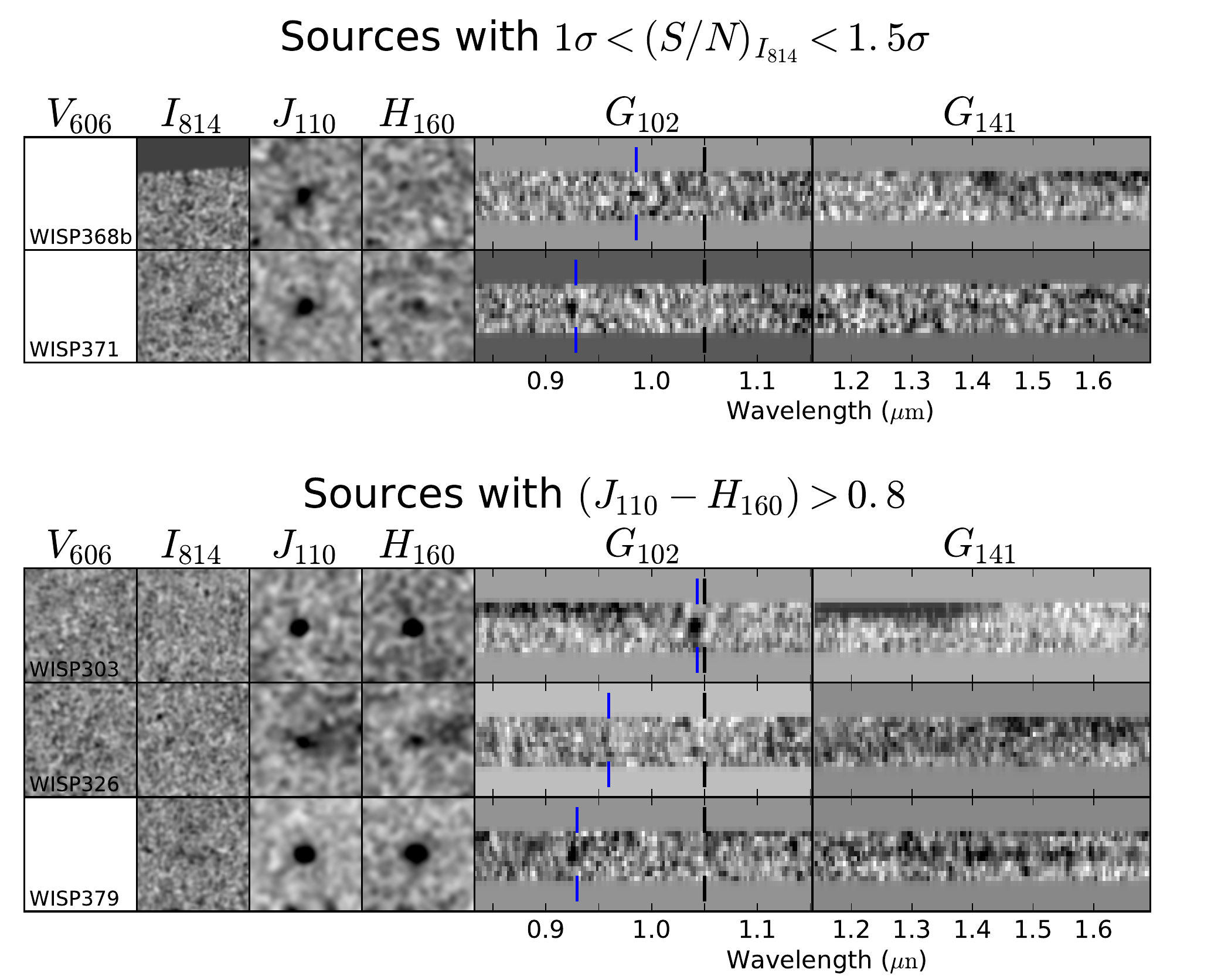}
\caption{Direct image postage stamps and the two-dimensional spectral stamps
of sources rejected from the LAE sample because of their flux in 
$\iband$ (\textit{top}) or their red $(\jband-\hband)$ colors 
(\textit{bottom}).
Columns show, from left to right, $\vband$, $\iband$, $\jband$, $\hband$,
$G_{102}$ and $G_{141}$.
The postage stamps are $3^{\prime\prime}$ on a side with the same
pixel scales and smoothing kernels as those of Figure \ref{fig:stamps}.
The grism spectral stamps have been smoothed by the same kernel as 
that of Figure \ref{fig:spectra}. In the $G_{102}$ stamps, the black lines
indicated the maximum wavelength cutoff of our criteria, 
$\lambda_{obs} = 1.05\mu$m. The blue lines indicate the wavelength of the 
emission line.
\label{fig:appplots}
}
\end{center}
\end{figure*}
In Figure \ref{fig:appplots}, we present the imaging and spectral stamps
of two categories of sources that 
were conservatively excluded from the LAE sample.
The top section shows sources that are marginally detected in $\iband$ 
at the $1\sigma$ level. These galaxies have 
$1\sigma < (S/N)_{\iband} < 1.5\sigma$ and are discussed in 
Section \ref{sec:iband}.
The bottom section shows sources that have IR colors that place them 
outside our color selection window: $(\jband-\hband) > 0.8$.
These galaxies may be very dusty or have strong $4000$-\AA\ breaks.
They are discussed in Section \ref{sec:others}.
Table \ref{tab:others} presents
the emission line fluxes and photometry for these five objects.

\begin{table*}
\begin{center}
\begin{threeparttable}
\caption{Additional Sources}
\label{tab:others}
\begin{tabular}{@{}lcccccccccccc}
\toprule
& ID & R.A. & Decl. & $f_{\mathrm{line}}$ & $\vband$ & $\iband$ & $\jband$ & $f_{J110}^{\mathrm{neb}} / f_{J110}^{\mathrm{total}}$ & $\hband$ \\
& & (J2000) & (J2000) & [$10^{-17}$ erg/s/cm$^2$] &  [mag] & [mag] & [mag] &  & [mag] \\
\midrule
& WISP368b & 23:22:35.32 & $-$34:50:41.6 & $6.1\pm 1.2$ & - & $27.6 \pm 0.7$ & $26.6\pm 0.2$ & $0.73$ & $27.18\pm 0.8$  \\
& WISP371 & 20:05:49.69 & $-$41:39:38.6 & $7.2\pm 1.1$ & - & $27.5\pm 0.6$ & $26.0\pm 0.1$ & $0.50$ & $25.8 \pm 0.3$  \\
\midrule
& WISP303 & 13:48:29.26 & $+$26:32:44.3 & $5.5\pm 0.6$ & $>28.26$ & $>27.35$ & $25.5\pm 0.1$ & $0.24$ & $24.8\pm 0.1$  \\
& WISP326 & 05:30:03.91 & $-$07:24:10.4 & $2.3\pm 0.6$ & $>28.09$ & $>27.55$ & $25.8\pm 0.3$ & $0.13$ & $25.7\pm 0.2$  \\
& WISP379 & 12:56:49.61 & $+$56:52:48.5 & $6.9\pm1.3$ & - & $>27.97$ & $25.5\pm0.2$ & $0.30$ & $24.3\pm0.2$  \\
\bottomrule 
\end{tabular}
\end{threeparttable}
\end{center}
\end{table*}

\end{document}